\documentclass[useAMS,usenatbib]{mn2e}

\usepackage{graphicx}
\usepackage{epsfig}
\usepackage{times}
\usepackage{natbib}
\usepackage{amsfonts}
\usepackage{amssymb}
\usepackage{amsmath}
\usepackage{amsfonts}

\title[Efficient cosmological parameter sampling using sparse grids] 
{Efficient cosmological parameter sampling using sparse grids} 
\author[M. Frommert, D. Pfl{\"u}ger, T. Riller, M. Reinecke, H.-J. Bungartz,
   T.~A. En{\ss}lin]
{M. Frommert$^{1,2}$\thanks{E-mail: mona@mpa-garching.mpg.de},
  D. Pfl{\"u}ger$^{1,3}$\thanks{E-mail: pflueged@in.tum.de},
  T. Riller$^{2}$, M. Reinecke$^{2}$, 
  H.-J. Bungartz$^{3}$, T.~A. En{\ss}lin$^{2}$ \\
$^{1}$ The authors contributed equally to the article\\
$^{2}$ Max-Planck-Institut f\"ur Astrophysik, Garching, Germany\\
$^{3}$ Institut f{\"u}r Informatik, TU M{\"u}nchen, Garching, Germany
}

\def\bm#1{{\hbox{\boldmath $#1$\unboldmath}}}
\def\vec#1{\textrm{\bm{#1}}}
\def\eg{{\rm e.g.$\,$}}
\def\ie{{\rm i.e.$\,$}}
\def\cf{{\rm cf.$\,$}}

\newcommand{\be}{\begin{equation}}
\newcommand{\ee}{\end{equation}}
\newcommand{\bea}{\begin{eqnarray}}
\newcommand{\eea}{\end{eqnarray}}

\newcommand{\Max}{{\rm{max}}}
\newcommand{\Ls}{{{\rm ls}}}
\newcommand{\km}{{\rm km}}
\newcommand{\Mpc}{{\rm Mpc}}

\newcommand{\bbN}{\mathbb{N}}

\newcommand{\bbR}{\mathbb{R}}

\newcommand{\vl}{\vec{l}}
\newcommand{\vi}{\vec{i}}
\newcommand{\vone}{\vec{1}}
\newcommand{\vx}{\vec{x}}
\newcommand{\vy}{\vec{y}}
\newcommand{\Order}{{\cal O}}
\newcommand{\LL}{\mathcal{L}}
\newcommand{\LLL}{\ln\mathcal{L}}
\newcommand{\LLLmax}{\ln\mathcal{L}_{\rm max}}
\DeclareMathOperator{\e}{e}
\newcommand{\Hide}[1]{}

\begin{document}

\date{Accepted ??? Received ???; in
  original form ???}
\pagerange{\pageref{firstpage}--\pageref{lastpage}} \pubyear{2008}
\maketitle
\label{firstpage}

\begin{abstract}

We present a novel method to significantly speed up cosmological
parameter sampling. The method relies on constructing an interpolation
of the CMB-log-likelihood based on sparse grids, which is used as
a shortcut for the likelihood-evaluation. We obtain excellent results
over a large region in parameter space, comprising about 25
log-likelihoods around the peak, and we reproduce the one-dimensional
projections of the likelihood almost perfectly. In speed and accuracy,
our technique is competitive to existing approaches to accelerate
parameter estimation based on polynomial interpolation or neural
networks, while having some advantages over them.  In our method,
there is no danger of creating unphysical wiggles as it can be the
case for polynomial fits of a high degree. Furthermore, we do not
require a long training time as for neural networks, but the
construction of the interpolation is determined by the time it takes
to evaluate the likelihood at the sampling points, which can be
parallelised to an arbitrary degree. Our approach is completely general,
and it can adaptively exploit the
properties of the underlying function. We can thus apply it to any
problem where an accurate interpolation of a function is needed.

\end{abstract}


%*****************************************************************************

\section{Introduction}

The main two bottlenecks in cosmological parameter estimation using
the power spectrum of the cosmic microwave background (CMB) are the
calculation of the theoretical $C_l$-spectrum using Boltzmann codes
such as {\small CMBFAST} \citep{cmbfast}, {\small CAMB} \citep{camb},
or {\small CMBEASY} \citep{cmbeasy} and the evaluation of the
likelihood using the official WMAP likelihood
code\footnote{http://lambda.gsfc.nasa.gov/product/map/dr3/likelihood\_get.cfm}.
There exist several methods to speed up the calculation of the power
spectrum \citep{cmbwarp, dash, habib} or the WMAP likelihood
function $\LL$ \citep{cmbfit, pico, cosmonet}. These methods are based
on different techniques, such as analytic approximations, polynomial
fits, and neural networks, which are all trained using a set of
training points, for which the real power spectra and likelihood
values have to be calculated.
Once the codes are trained for a particular cosmological model, they
can be used to evaluate the power spectrum or the likelihood function
in every subsequent parameter estimation run, which significantly
speeds up the Markov Chain Monte Carlo (MCMC) simulations used for
parameter estimation.  
Due to the ever-growing amount of available data, a fast evaluation of
the likelihood is becoming of increasing importance, especially when
combining CMB data with data-sets whose likelihood is less expensive to
evaluate. The {\it Planck Surveyor} mission \citep{planck} will be the upcoming
challenge in this respect. 

In this work, we approximate the WMAP log-likelihood function $\LLL$
in the spirit of CMBfit \citep{cmbfit} and Pico \citep{pico}, which
work with polynomial fits, and CosmoNet \citep{cosmonet},
an approach based on neural networks. In
contrast to the fitting functions constructed therein, we introduce
the technique of \emph{sparse grids} in this context to construct an
interpolation of $\LLL$, returning the exact function values at a set of
sampling points.

Most straightforward interpolation techniques are based on sets of
sampling points in each dimension, typically based on (uniform) grid
structures---consider, e.\,g., piecewise $d$-linear or piecewise
$d$-polynomial interpolation schemes. Unfortunately, grid-based methods
are only feasible in low-dimensional settings, as they suffer from the
so-called \emph{curse of dimensionality}: Spending $\tilde{N}$
function evaluations or grid points in one dimension leads to
$\tilde{N}^d$ grid points in $d$ dimensions. The exponential
dependency on the dimensionality imposes severe restrictions on the
number of dimensions that can be handled. Sparse grids, as introduced
by \cite{zenger91sparse}, allow to overcome the curse
of dimensionality to some extent, at least for sufficiently smooth
functions as it is the case in our setting. Sparse grid interpolation
is based on an a priori selection of grid points, requiring
significantly fewer grid points than conventional interpolation on a
full grid, while preserving the asymptotic error decay of a full-grid
interpolation with increasing grid resolution up to a logarithmic
factor. This permits us to compute higher-dimensional interpolations and
approximations than before. A very good overview about sparse grids,
discussing general properties, can be found in \cite{bungartz04sparse}.

The sparse grid technique is a completely general approach, not
tailored to a single application, and can therefore be used to
interpolate any function which is sufficiently smooth.
Additionally, as it allows for arbitrary adaptive refinement
schemes, the general, fast convergence rates can be improved even
further, by adapting to the special characteristics of the underlying
target function.

We obtain excellent results, which are competitive to fitting
procedures using polynomials \citep{pico,cmbfit} or neural networks
\citep{cosmonet} in speed and accuracy. Furthermore, we believe that
the interpolation based on sparse grids has several advantages over these
approaches. First of all, we can use the results of sparse grid
approximation quality \citep{bungartz04sparse},
guaranteeing the convergence of the interpolation towards the original
function with increasing number of grid points.

Second, once we have chosen the volume in which we want to interpolate
the function in question, the sparse grid structure itself determines
a priori the location of potential sampling points (which can
additionally be refined in an adaptive manner a posteriori). This
makes it unnecessary to assemble a set of training points
beforehand (e.g. by running MCMCs as it is done by \cite{pico}). The
generation of the sampling points and the construction of the
interpolant can be strongly parallelised, which makes the sparse grid
approach an ideal candidate for computational grid projects such as
the
AstroGrid\footnote{http://www.d-grid.de/index.php?id=45\&L=1}.
The time needed to construct the interpolant is determined almost only
by the time it takes to evaluate the likelihood at the sampling
points. We do not need additional training time as in the case of
\cite{cosmonet}.

Furthermore, polynomial fits to a set of training points run the risk
of creating unphysical wiggles if the polynomial degree of the fitting
function is chosen too high with respect to the amount of training
points available. Using the sparse grids approach, piecewise polynomials of low
degree are sufficient, as we are not fitting certain evaluation
points, but rather interpolating a function. Increasing the accuracy
is therefore equivalent to evaluating at more sampling points.

Sparse grids are based on a hierarchical formulation of the underlying
basis functions, which can be used to obtain a generic estimate of the
current approximation error while evaluating more and more sampling
points. This can be directly used as a criterion for adaptive
refinement as well as to stop further refinement.

Another advantage is that the projection of sparse grid
interpolations can be done in a very fast and simple way.
This would make sparse grids in principle a good candidate for
sampling posteriors and projecting them directly, without having to
use a Markov Chain approach in order to marginalise the
posterior. Given that MCMCs need to determine the points
 sequentially and can therefore
not be parallelised (apart from running several chains at the same
time), it would be highly desirable to find alternatives that can be
run in parallel.

We have attempted to use sparse grids in order to substitute the MCMCs in
cosmological parameter estimations. In
order to directly project the posterior distribution we would need to 
sample the posterior rather than its logarithm. Since, in general, the
logarithm of a probability density function is considerably more well-behaved
than the function itself, \cite{cmbfit}, \cite{pico}, and
\cite{cosmonet} all operate in log-space to speed up the generation of
MCMCs instead. 
As the convergent phase of the interpolation with sparse grids sets in
rather late when interpolating Gaussian functions (and thus 
the WMAP likelihood, which is close to
a $d$-dimensional Gaussian), we restricted ourselves to the
log-likelihood and thus to accelerating MCMCs, too.

The article is organised as follows. First, we describe the basics of
sparse grids in Sec.\ \ref{sec:sg_basics}, introducing a modification
of the standard sparse grid approach, thus adapting the latter to our
problem. In Sec.\ \ref{sec:sg_wmap}, we then present the interpolation
of the WMAP likelihood for two different sets of parameters in both
six and seven dimensions. We show that the results obtained for
regular (non-adaptive) sparse grids are already competitive to other
approaches and demonstrate how adaptive refinement can further improve 
the results. Sec.\ \ref{sec:conclusions} finally concludes our work.

\section{Basics of sparse grids}\label{sec:sg_basics}
\label{sec:basics}

Standard grid-based approaches of interpolating a function $f$ exhibit
the curse of dimensionality, a term going back to 
\cite{bellman61adaptive}: Any straightforward discretisation scheme which
employs $\tilde{N}$ grid points (or, equivalently, degrees of freedom) 
in one dimension leads to $\tilde{N}^d$ grid points in $d$
dimensions. For reasonable $\tilde{N}$, the exponential dependency on
the number of dimensions typically does not allow to handle more than
four-dimensional problems.

Sparse grids are able to overcome this hurdle to some extent,
requiring significantly fewer grid points than a full grid, while
preserving the asymptotic error decay of full grid interpolation with
increasing grid resolution up to a logarithmic
factor. Sparse grids have originally been developed for the solution
of partial differential equations \citep{zenger91sparse} and have
meanwhile been applied to various problems, see
\cite{bungartz04sparse} and the references cited therein. Recent work
on sparse grids includes stochastic and non-stochastic partial
differential equations in various settings
\citep{petersdorff06sparse,ganapathysubramanian07sparse,widmer08sparse},
as well as applications in economics
\citep{reisinger07efficient,holtz08sparse}, regression
\citep{garcke09fitting,garcke06regression}, classification
\citep{bungartz08adaptive,garcke01data}, fuzzy modelling
\citep{klimke06fuzzy}, and more. Note that (non-adaptive) sparse grids
are closely related to the technique of hyperbolic crosses
\citep{temlyakov93approximation}.

In this section, we provide a brief overview of sparse grids for
interpolation. For a detailed derivation of the characteristics of
sparse grids, we refer to \cite{bungartz04sparse}.
We start by formulating the interpolation on a conventional full
grid using hierarchical basis functions, from which we
then derive the interpolation on a sparse grid by omitting the
basis functions contributing least to the interpolation.

\subsection{General idea of interpolation on a full grid}

We consider the piecewise $d$-linear interpolation of a function
$f:\nolinebreak 
\Omega \rightarrow \bbR$ which is given only algorithmically, i.e., we
have no closed form of $f$ but we can only evaluate $f$ at arbitrary
points using a numerical code. As we want to discretise our domain of
interest $\Omega$, we restrict $\Omega$ to a compact sub-volume of
$\bbR^d$; here, $\Omega \equiv [0,1]^d$, the $d$-dimensional
unit-hypercube. (For the standard approach of sparse grids techniques,
we only consider functions that are zero on the boundary of the volume
on which they are defined. This assumption will be dropped when we
come to the interpolation of the log-likelihood of WMAP.)

\begin{figure}
\begin{center}
\includegraphics[width=0.45\textwidth]{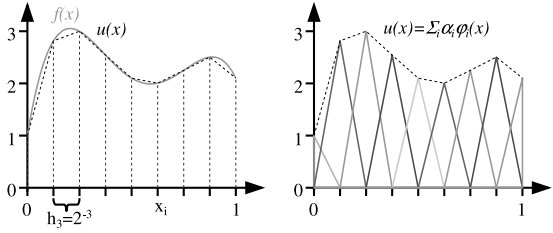}
\end{center}
\caption[One-dimensional piecewise linear
  interpolation]{One-dimensional piecewise linear interpolation
  $u(x)$, dashed, 
  of a function $f(x)$, solid, (left) by a
  linear combination of hat basis functions (right).}
\label{fig:1d_interpol}
\end{figure}

To construct an interpolant $u$ of $f$, we discretise $\Omega$ via a
regular grid, obtaining equidistant grid points $\vx_i$, with mesh width
$h_n=2^{-n}$ for some discretisation or refinement level $n$, at which
we evaluate and interpolate $f$. If we define a suitable set of
piecewise $d$-linear basis functions $\varphi_i(\vx)$, we can obtain
$u(\vx)$ from the space of continuous, piecewise $d$-linear functions
$V_n$ by combining them adequately as a weighted sum of basis
functions, i.e.\
\[
  f(\vx) \approx u(\vx) \equiv \sum_{i} \alpha_{i}
  \varphi_{i}(\vx)
\]
with coefficients $\alpha_i$. Figure \ref{fig:1d_interpol} sketches
the idea for a one-dimensional example, using the standard nodal
basis.

The curse of dimensionality, encountered when using a full grid, can be
circumvented by a suitable choice of basis functions: We need a basis
where the relevant information is represented by as few basis
functions as possible. Most basis
functions can then be omitted as they contribute only little to the
interpolation of $f$, reducing a full grid to a sparse grid and
allowing us to handle higher-dimensional functions than before. A
suitable basis can be found by a hierarchical construction as
introduced in the following section.

\subsection{Hierarchical basis functions in one dimension}

\begin{figure}
\begin{center}
\includegraphics[width=0.45\textwidth]{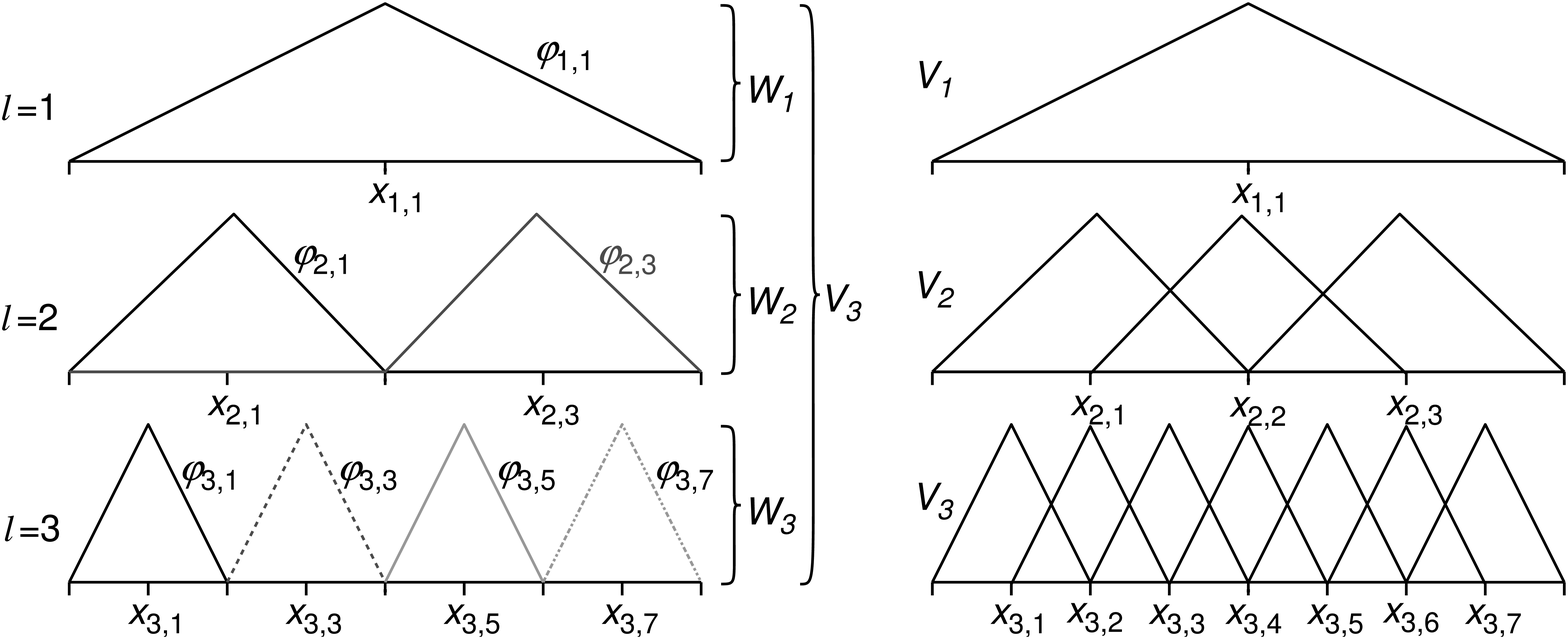}
\end{center}
\caption[One-dimensional basis functions]{One-dimensional basis
  functions $\varphi_{l,i}$ and 
  corresponding grid points $x_{l,i}$ up to level $n=3$ in the
  hierarchical basis (left) and the common nodal point basis (right).}
\label{fig:basis-fcns}
\end{figure}

Sparse grids depend on a hierarchical decomposition of the underlying
approximation spaces. Therefore, and first considering only the one-dimensional case
which we will later extend to $d$ dimensions, we use the standard hat function,
\begin{equation}
  \varphi(x) = \max( 1-|x|, 0)\,,
\end{equation}
from which we derive one-dimensional hat basis functions by dilatation
and translation,
\begin{equation}
  \varphi_{l,i}(x) \equiv \varphi(2^l x -i)\,,
\end{equation}
which depend on a level $l$ and an index $i$, $0<i<2^l$. The basis
functions have local support and are centred at grid points
$x_{l,i}=2^{-l}i$, at which we will interpolate $f$. Introducing the
hierarchical index sets
\begin{equation}
  I_l \equiv \left\{\, i \in \bbN: 1\leq i\leq2^l-1, i \text{~odd}\,\right\}\,,
\end{equation}
we obtain a set of hierarchical subspaces $W_l$,
\begin{equation}
  W_l \equiv \textrm{span}\left\{\varphi_{l,i}(x) : i \in I_l\right\}.
\end{equation}
We can then formulate the space of piecewise linear functions $V_n$ on
a full grid with mesh width $h_n$ for a given level $n$ as a direct
sum of $W_l$,
\begin{equation}
  V_n = \bigoplus_{l \leq n} W_{l}\,,
\end{equation}
see Figure \ref{fig:basis-fcns}. Note that all basis functions of the
same subspace $W_l$ have the same size, shape, and compact support,
that their supports are non-overlapping, and that together they cover the whole domain.

The interpolation $u(x)\in V_n$ can then be written as a finite sum,
\begin{equation}
  u(x) = \sum_{l \leq n, i\in I_l} \alpha_{l,i} \varphi_{l,i}(x)\,,
\end{equation}
where the so-called (hierarchical) surpluses $\alpha_{l,i}$ are uniquely
indexed by the same level and index as the corresponding basis
functions.

\subsection{Higher-dimensional interpolation on a full grid}

\begin{figure}
\begin{center}
\begin{minipage}[t]{0.4\textwidth}
\small
\centering
\begin{minipage}[t]{0.29\textwidth}
\centering %
\includegraphics[width=0.9\textwidth]{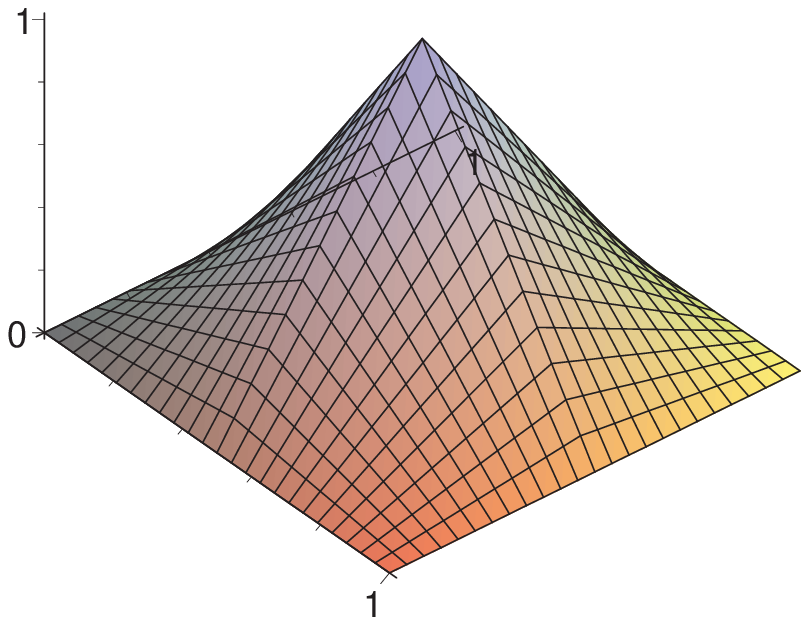}\\
$W_{(1,1)^T}$ %
\end{minipage}
\begin{minipage}[t]{0.29\textwidth}
\centering %
\includegraphics[width=0.9\textwidth]{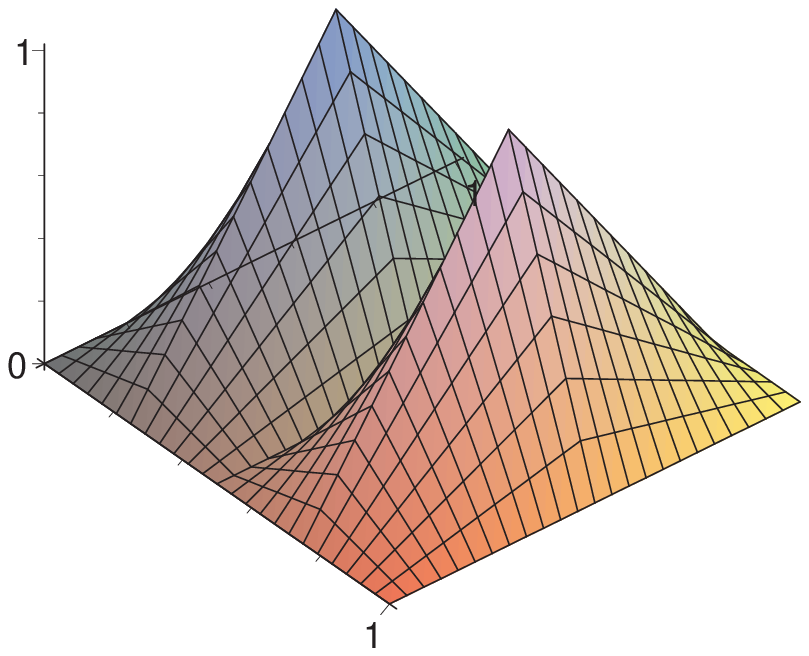}\\
$W_{(2,1)^T}$ %
\end{minipage}
\begin{minipage}[t]{0.29\textwidth}
\centering %
\includegraphics[width=0.9\textwidth]{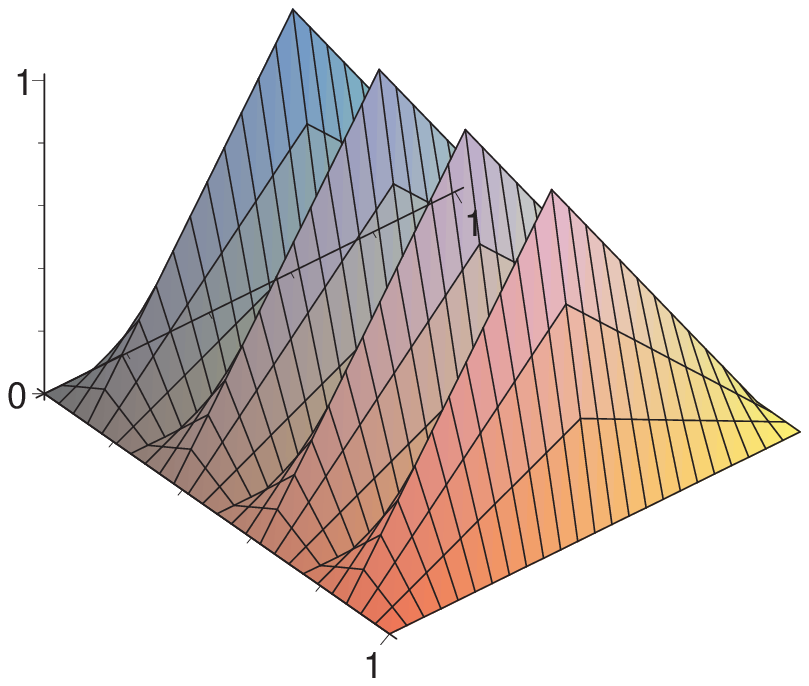}\\
$W_{(3,1)^T}$ %
\end{minipage} 
\begin{minipage}[t]{0.08\textwidth}
\vspace{-3em}$\cdots$
\end{minipage} \\[1em]
\begin{minipage}[t]{0.29\textwidth}
\centering %
\includegraphics[width=0.9\textwidth]{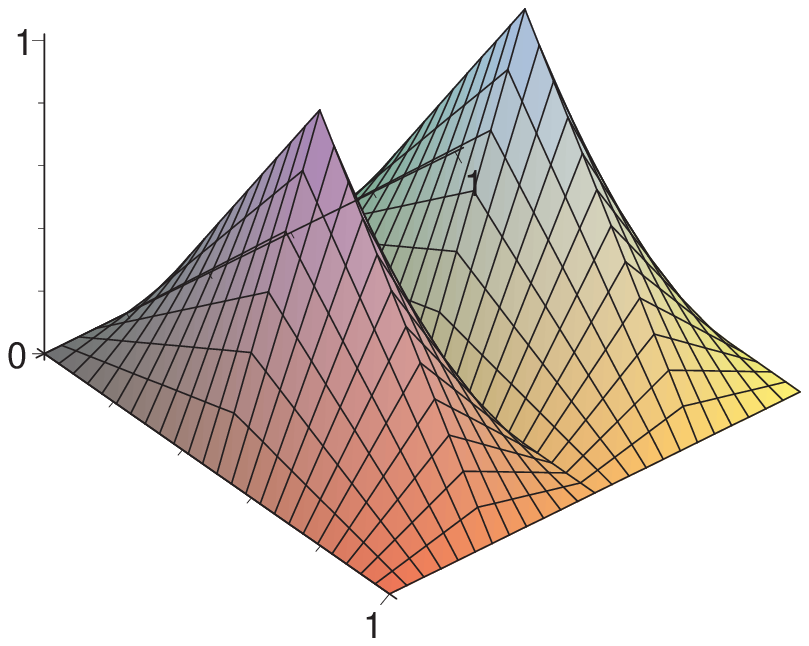}\\
$W_{(1,2)^T}$ %
\end{minipage}
%\hspace{1cm}
\begin{minipage}[t]{0.29\textwidth}
\centering %
\includegraphics[width=0.9\textwidth]{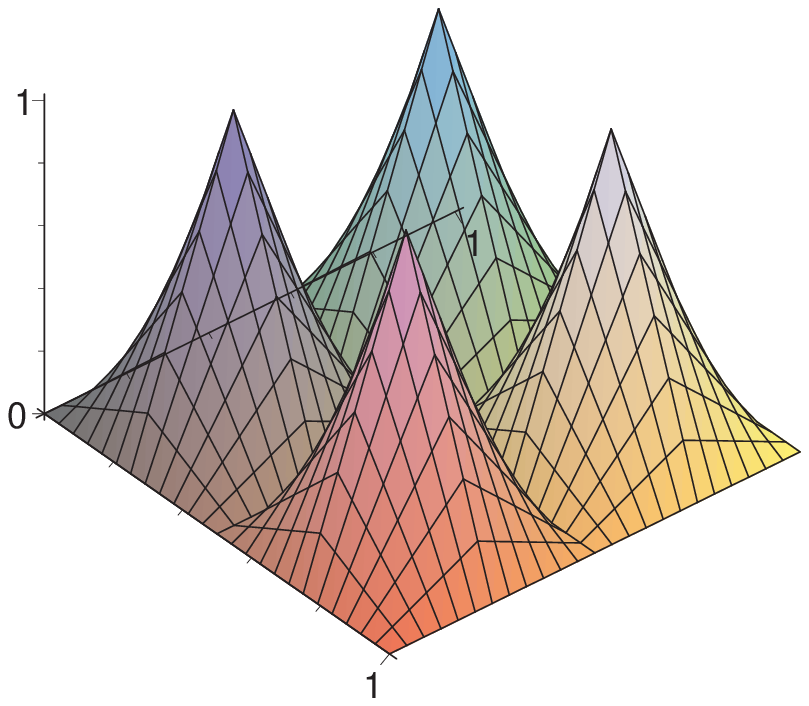}\\
$W_{(2,2)^T}$ %
\end{minipage}
%\hspace{1cm}
\begin{minipage}[t]{0.29\textwidth}
\centering %
\includegraphics[width=0.9\textwidth]{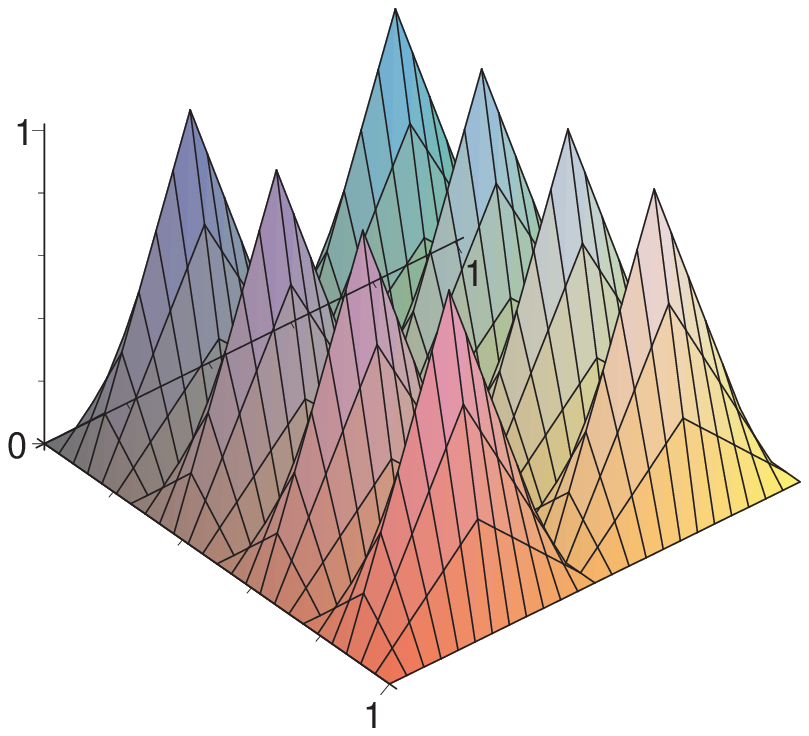}\\
$W_{(3,2)^T}$ %
\end{minipage} 
\begin{minipage}[t]{0.1\textwidth}
\vspace{-3em}$\cdots$
\end{minipage} \\[1em]
\begin{minipage}[t]{0.29\textwidth}
\centering %
\includegraphics[width=0.9\textwidth]{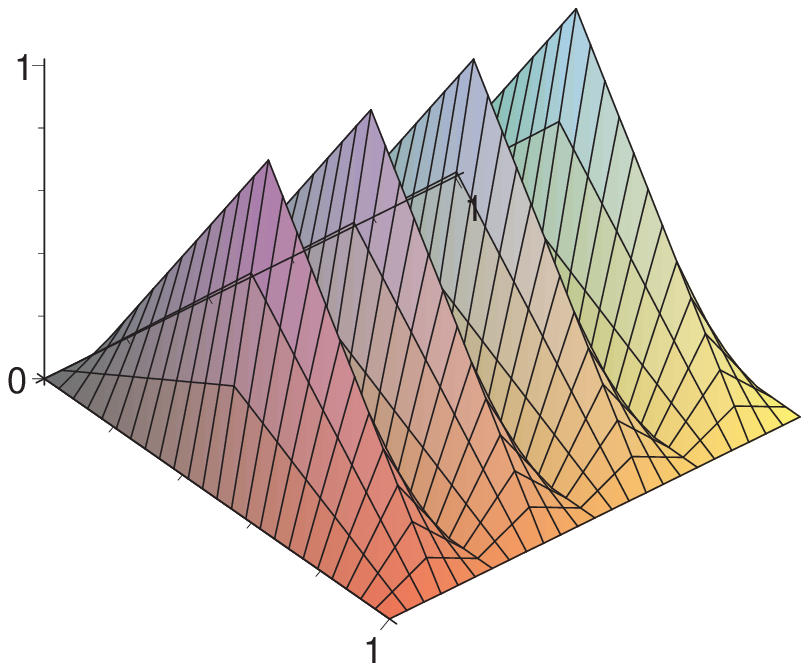}\\
$W_{(1,3)^T}$ %
\end{minipage}
%\hspace{1cm}
\begin{minipage}[t]{0.29\textwidth}
\centering %
\includegraphics[width=0.9\textwidth]{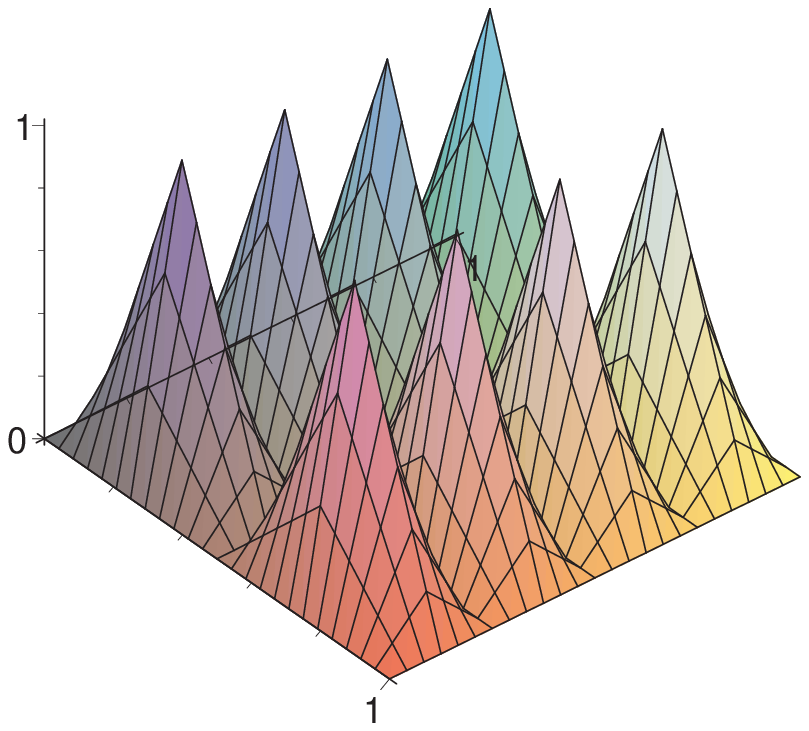}\\
$W_{(2,3)^T}$ %
\end{minipage}
%\hspace{1cm}
\begin{minipage}[t]{0.29\textwidth}
\centering %
\includegraphics[width=0.9\textwidth]{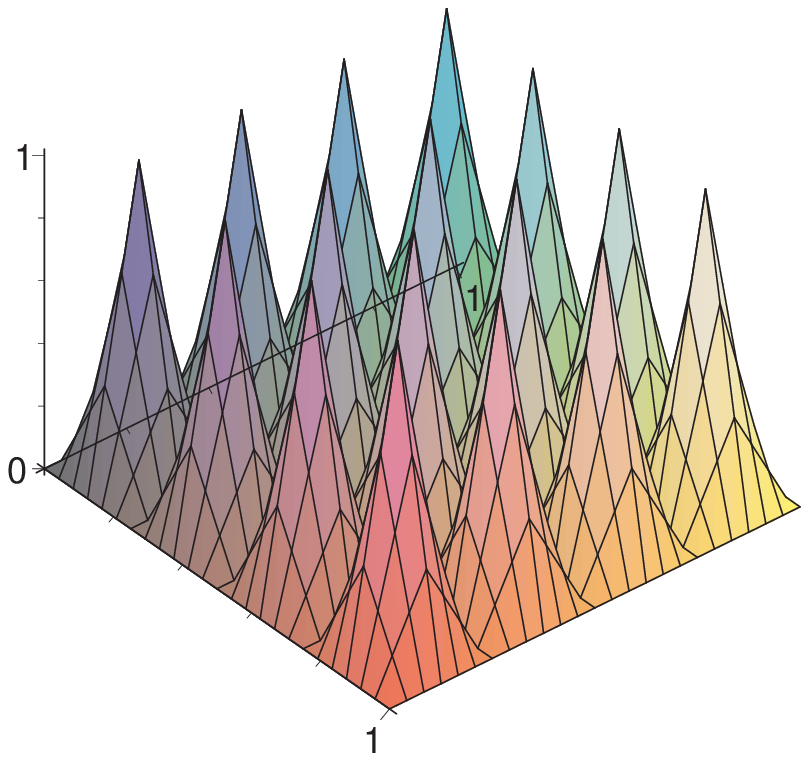}\\
$W_{(3,3)^T}$ %
\end{minipage} 
\begin{minipage}[t]{0.1\textwidth}
\vspace{-3em}$\cdots$
\end{minipage} \\
\begin{minipage}[t]{0.29\textwidth}
\centering$\vdots$
\end{minipage}
%\hspace{0.5cm}
\begin{minipage}[t]{0.29\textwidth}
\centering$\vdots$
\end{minipage}
%\hspace{0.5cm}
\begin{minipage}[t]{0.29\textwidth}
\centering$\vdots$
\end{minipage}
\begin{minipage}[t]{0.1\textwidth}
$\ddots$
\end{minipage}
\end{minipage}
\end{center}
\caption[Two-dimensional basis functions]{Basis functions of the
  subspaces $W_\vl$ for 
  $|\vl|_\infty\leq 3$ in two dimensions.}
\label{fig:2d-basis-fcns}
\end{figure}

The basis functions are extended to the $d$-dimensional case
via a tensor product approach,
\begin{equation}
  \varphi_{\vl,\vi}(\vx) \equiv \prod_{j=1}^d
  \varphi_{l_j,i_j}(x_j)\,,
\end{equation}
with the $d$-dimensional multi-indices $\vl$ and $\vi$ indicating
level and index for each dimension. The other one-dimensional notations
can be transferred to the arbitrary-dimensional case as well,
consider, \eg, the index set $I_\vl$,
\begin{equation}
  I_\vl \equiv \left\{\, \vi: 1 \leq i_j\leq2^{l_j}-1, i_j \text{~odd}, 1 \leq
  j \leq d\,\right\}\,,
\end{equation}
the subspaces $W_{\vl}$, the space $V_n$ of piecewise $d$-linear functions
with mesh width $h_n$ in each dimension,
\begin{equation}
  V_n = \bigoplus_{|\vl|_\infty \leq n} W_{\vl}\,,
\end{equation}
leading to a full grid with $(2^n-1)^d$ grid points, and to the
interpolant $u(\vx)\in V_n$,
\begin{equation}
  u(\vx) = \sum_{|\vl|_\infty \leq n, \vi\in I_\vl} \alpha_{\vl,\vi} \varphi_{\vl,\vi}(\vx)\,.
\label{eq:interpol_full_grid}
\end{equation}
Here and later on, we need the $l_1$-norm $|\vl|_1=\sum_{j=1}^d l_j$
and the maximum-norm $|\vl|_\infty=\max_{1\leq j \leq d} l_j$ of
multi-indices $\vl$. Figure \ref{fig:2d-basis-fcns} shows some
2-dimensional examples for the basis functions of the subspaces
$W_\vl$, which correspond to anisotropic sub-grids with mesh-width
$h_{l_j}$ in dimension $j$ characterised by the multi-index $\vl$.

\subsection{Sparse grids}

Starting from the hierarchical representation of $V_n$ by the
subspaces $W_{\vl}$, we can now select those subspaces that contribute
most to the overall solution of the full-grid interpolation
(\ref{eq:interpol_full_grid}). If the function we want to approximate
meets certain smoothness conditions---the mixed second derivatives
have to be bounded---this can be done a priori as we can derive bounds
for the contributions of the different subspaces. We then obtain
the sparse grid space
\begin{equation}
  V_n^{(1)} \equiv \bigoplus_{|\vl|_1 \leq n+d-1} W_{\vl}\,,
\label{eq:vn1}
\end{equation}
leaving out those subspaces from the full grid space $V_n$ with many
basis functions of small support. (The exact choice of subspaces
depends on the norm in which we measure the error; the result above is
optimal for both the $L_2$ norm and the maximum norm.) Note that in
the one-dimensional case, the sparse grid space equals the full grid
space.

\begin{figure}
\begin{center}
\begin{minipage}[t]{0.6\textwidth}%
\vspace{0cm}%
\hspace{-0.1\textwidth}
\begin{minipage}[t]{0.8\textwidth}%
\vspace{0cm}%
\begin{center}
\includegraphics[width=0.6\textwidth]{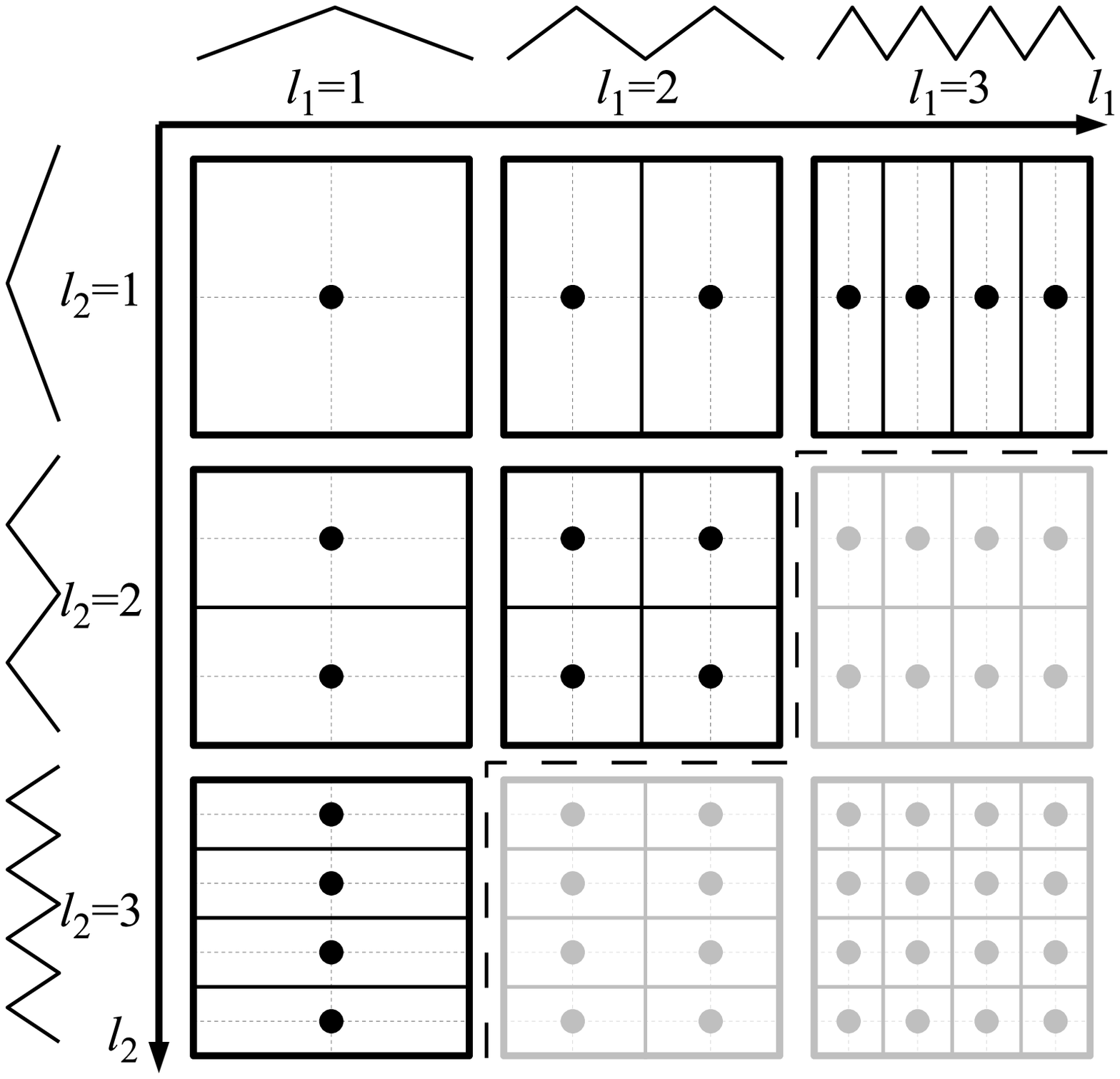}
\end{center}
\end{minipage}%
\hspace{-0.1\textwidth}
\begin{minipage}[t]{0.31\textwidth}%
\vspace{0cm}%
\centering
\vspace{0.65\textwidth}
\hspace{-0.5\textwidth}
\includegraphics[width=0.45\textwidth]{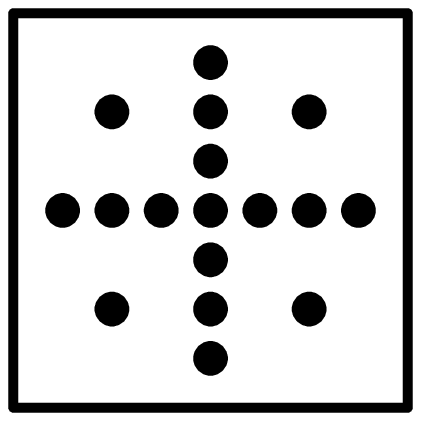}\\
\hspace{-0.5\textwidth}
$V^{(1)}_3$
\end{minipage}%
\end{minipage}%
\end{center}
\caption[Two-dimensional sub-grids of a sparse grid]{The
  two-dimensional subspaces $W_{\vl}$ up to $l=3$ 
  ($h_3=1/8$) in each dimension. The optimal selection of subspaces
  (black) and the corresponding sparse grid on level $n=3$ for the
  sparse grid space $V_3^{(1)}$. The corresponding full grid of level
  3 corresponds to the direct sum of all subspaces that are shown.}
\label{fig:subgrids}
\end{figure}

\begin{figure}
\begin{center}
\includegraphics[width=0.2\textwidth]{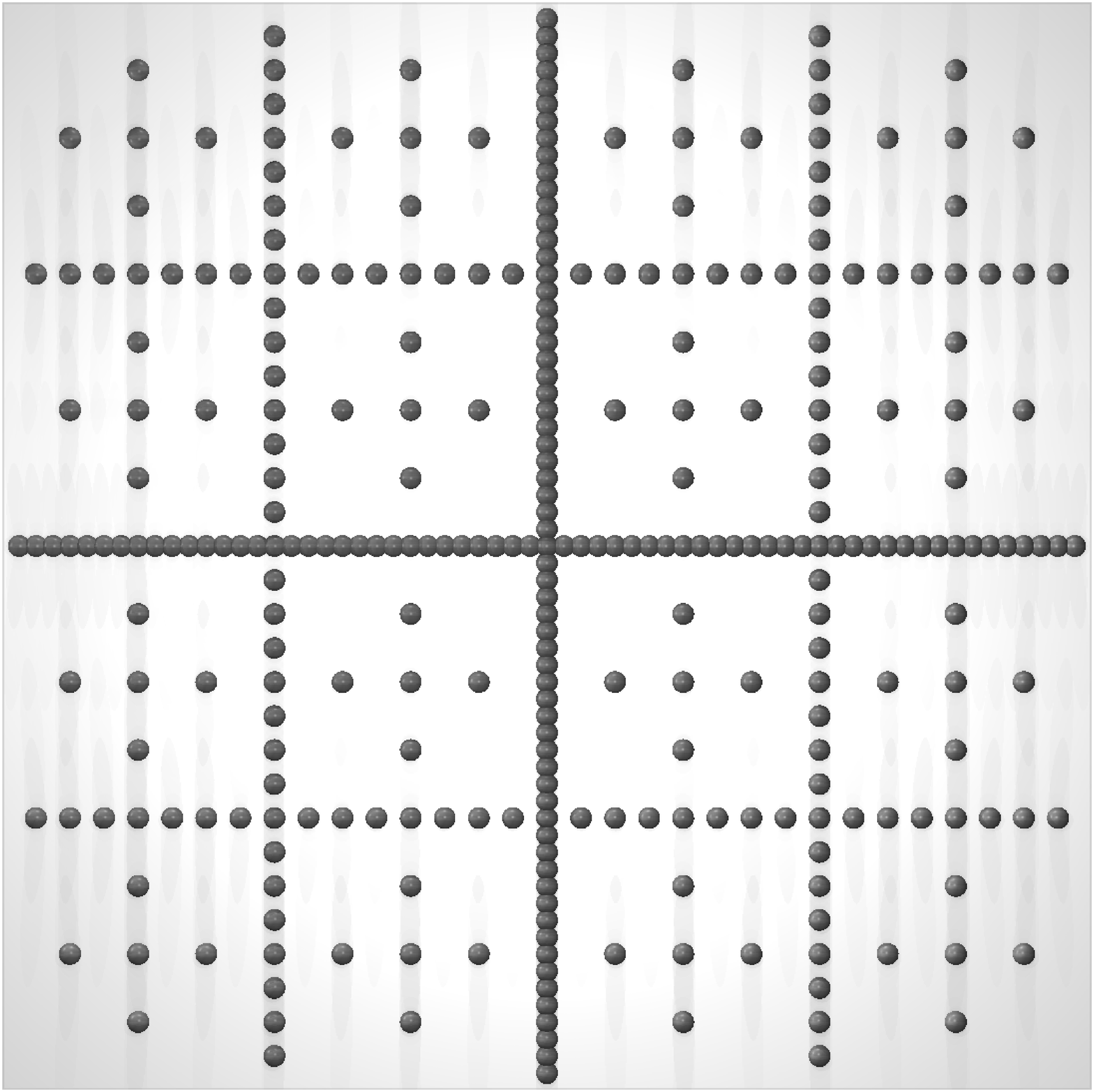}
\includegraphics[width=0.2\textwidth]{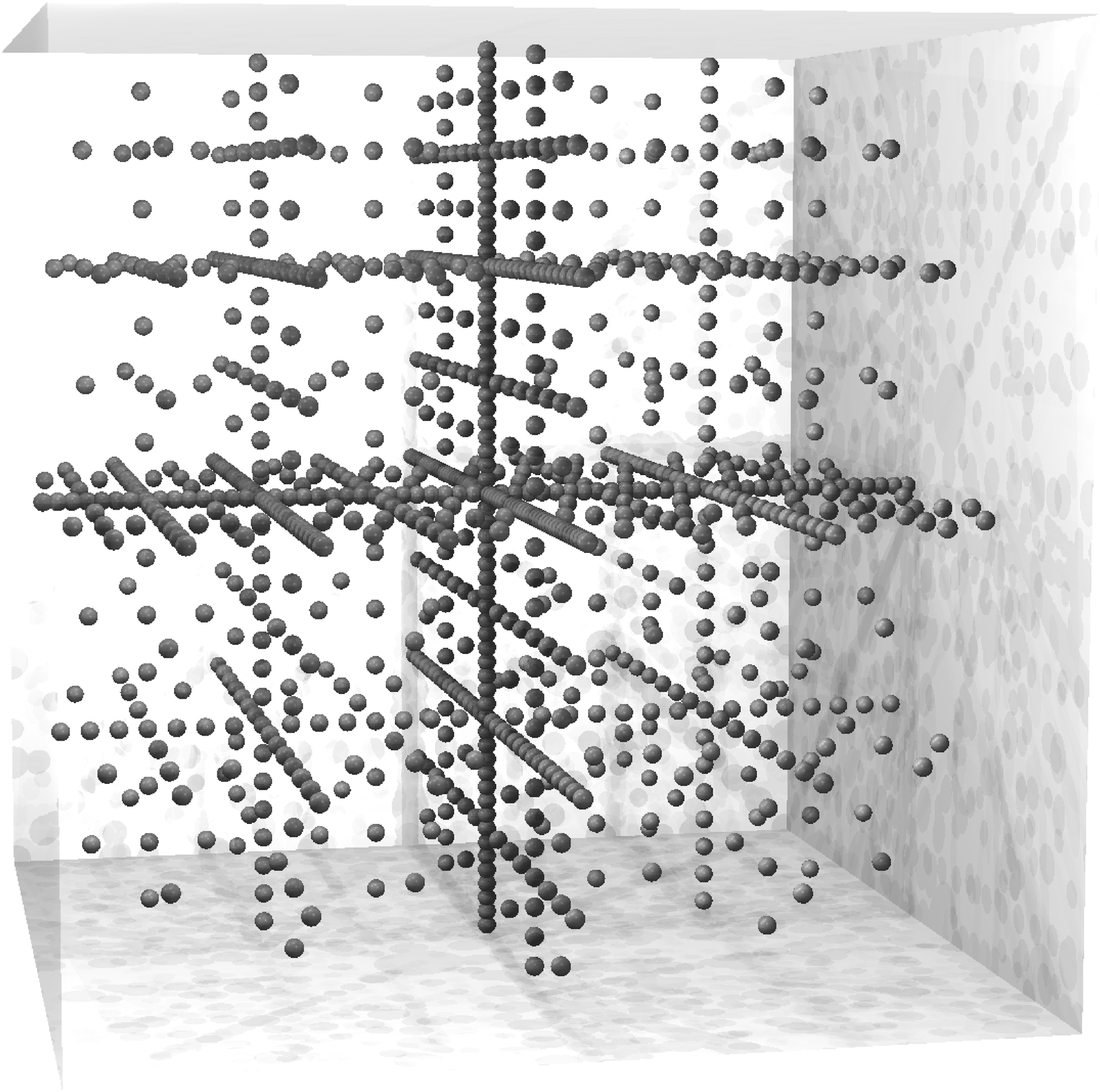}
\caption[Sparse grids in two and three dimensions]{Sparse grids in two
  and three dimensions for level $n=6$.} 
\label{fig:sg_2d_3d}
\end{center}
\end{figure}

Figure
\ref{fig:subgrids} shows the selection of subspaces and the resulting
sparse grid for $n=3$, i.e.\ the sparse grid space
$V_3^{(1)}$. Compared to the full grid for the same discretisation
level $n$ (the full grid space $V_3$ would also comprise the grey
subspaces in Figure \ref{fig:subgrids}), this
reduces the number of grid points (and therefore function evaluations
and unknowns) significantly from $\Order(h_n^{-d})=\Order(2^{nd})$ to
$\Order(h_n^{-1} (\log h_n^{-1})^{d-1})$ -- whereas the asymptotic
accuracy deteriorates only slightly from $\Order(h_n^2)$ to
$\Order(h_n^2 (\log h_n^{-1})^{d-1})$, see \cite{bungartz04sparse} for
detailed derivations. Figure \ref{fig:sg_2d_3d} shows sparse grids in
two and three dimensions for level $n=6$ each.

Functions which do not meet the smoothness requirements or which show
significantly differing characteristics (comprising steep regions as
well as flat ones, e.g.) can be tackled as well, if adaptive
refinement is used. The sparse grid structure (\ref{eq:vn1}) defines
an a priori selection of grid points which is optimal if certain
smoothness conditions are met and no further knowledge about the
function in question is known or used. An adaptive (a posteriori)
refinement can additionally select which grid points in a sparse grid
structure should be refined next, due to local error estimation,
e.g. To refine a grid point, often all $2d$ children in the
hierarchical structure are added to the current grid, if they haven't
been created yet. Note that it usually has to be ensured that all
missing parents have to be created, as algorithms working on sparse
grids depend on traversals of the hierarchical tree of basis
functions. If additional knowledge about the problem at hand is
available, it can be used in the criterion for adaptive refinement,
allowing to better adapt to problem specific characteristics.

\subsection{Extension to functions that are non-zero on the boundary}

Up until now we have only considered functions that are zero on the
domain's boundary $\delta\Omega$. To allow for non-zero values on the
boundary, usually additional grid points located directly on
$\delta\Omega$ are introduced. For example, the one-dimensional basis
on level one, containing only $\varphi_{1,1}(x)$, is extended by two
basis functions with level $0$ and indices $0$ and $1$ restricted to
$\Omega$, namely $\varphi_{0,0}(x)$ and $\varphi_{0,1}(x)$. They are
then extended to the $d$-dimensional case as before, with the
exception that the new basis now contains basis functions on the
modified level one with overlapping support.

Apparently, this approach results in many more grid points (and
therefore expensive function evaluations) than before. This shows
quite nicely that it is not sufficient to just consider the
asymptotic behaviour: asymptotically, nothing changes, but for very
high dimensionalities we are not able to even start to interpolate any
more. In $d$ dimensions, the basis for the subspace $W_\vone$ for
example contains already $3^d$ basis functions, rather than a single
one. Especially in settings where a very high accuracy close to the
boundary is not required---which holds in our case---(or where an
adaptive selection of grid points is used in any case), it can be
advantageous to omit the grid points on the boundary, and instead
modify the basis functions to extrapolate towards the boundary of the
domain.

\begin{figure}
\begin{center}
\includegraphics[width=0.18\textwidth]{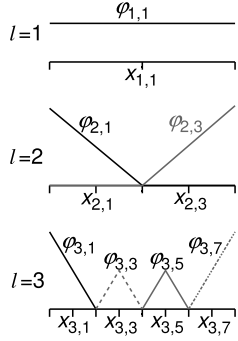}
\end{center}
\caption[Modified one-dimensional basis functions]{Modified
  one-dimensional basis functions $\varphi_{l,i}$: 
  constant on level 1 and ``folded up'' if adjacent to the boundary on
  all other levels.}
\label{fig:basisfunktionen_ausgeklappt}
\end{figure}

We modify the one-dimensional basis functions as
follows: On level 1, we have only one degree of freedom; the best guess
towards the boundary is to assume the same value, leading to a
constant basis function. On all other levels, we extrapolate
linearly towards the boundary, ``folding up'' the uttermost basis
functions. All other basis functions remain unchanged, yielding 
\begin{equation}
\varphi_{l,i}(x) \equiv \left\{ \begin{array}{ll}
    1 & \mbox{if } l = 1 \wedge i = 1 \,,\\[4pt]
    \varphi_{l,i}^{\rm left}(x) & \mbox{if } l > 1 \wedge i=1\,,\\[4pt]
     \varphi_{l,i}^{\rm right}(x) & \mbox{if } l > 1 \wedge i=2^l-1\,,\\[4pt]
    \varphi\left(x \cdot 2^l - i\right) & \mbox{else}\,,
\end{array}
\right.
\label{eqn:phi_li2}
\end{equation}
with
\bea
\varphi_{l,i}^{\rm left}(x) \!\!&\equiv&\!\! \left\{ 
      \begin{array}{ll} 
      2-2^l \cdot x ~~~~~& \mbox{if } x \in \left[0,\frac{1}{2^{l-1}}\right]\\
      0 & \mbox{else}
      \end{array} 
    \right\} \,, \\
\varphi_{l,i}^{\rm right}(x) \!\!&\equiv&\!\! \left\{ 
      \begin{array}{ll} 
      2^l \cdot x+1-i & \mbox{if } x \in \left[1-\frac{1}{2^{l-1}},1\right]\\
      0 & \mbox{else}
      \end{array}
    \right\} \,.
\eea
Examples of the modified one-dimensional basis functions
are shown in Figure \ref{fig:basisfunktionen_ausgeklappt}. 
The $d$-dimensional basis functions are obtained as before via a tensor
product of the one-dimensional ones.

\section{Interpolation of the WMAP likelihood surface}
\label{sec:sg_wmap}

We now construct an interpolation of the WMAP
log-likelihood, $\LLL$, using sparse grids.  
In order to adapt the problem to our interpolation approach, 
we first use a 6-dimensional set of so-called normal parameters
introduced in \cite{cmbfit}, which
are a transformation of the usual cosmological parameters such that the
major axes of the Gaussian align with the coordinate axes. The
logarithm of the likelihood is then well-approximated by a sum of 
one-dimensional parabolas
in the different parameters, a fact that we will take advantage of by using
the modified basis-functions described in
(\ref{eqn:phi_li2}). For this set of normal 
parameters, we obtain an accurate interpolation already for a
comparably low refinement level. This is shown for the 6-dimensional
model as well as for a 7-dimensional extension, using the running of
the spectral index as an additional parameter. 

However, as a subsequent step we demonstrate that the parameter
transformation is not essential for obtaining a good interpolation. By
investing more grid points, we obtain an accurate interpolation as well
when using directly the 6- and 7-dimensional standard parameter
set, which is usually used in cosmological parameter sampling.
This approach shows the advantage of sparse grids of being rather generic.
Furthermore, we are not restricted to the parameter range in
which the transformation to normal parameters can be inverted.

\subsection{Choice of basis functions}

We use the modified basis functions as introduced in
(\ref{eqn:phi_li2}), which are well-suited for our problem.
First, and as already indicated in Sec.\ \ref{sec:basics}, the region
close to the domain's boundary is less important in our setting than
the centre of $\Omega$: We will centre the domain of interest roughly
at the maximum of the log-likelihood function $\LLL$ and determine the
boundary such that it includes the region with
$(\LLLmax-\LLL)\lesssim 25$, which we will refer to as the 25
log-likelihood region (see Sec.\ \ref{sec:test_set}).

Towards the boundaries of our intervals,
the likelihood is then effectively zero and thus no great accuracy is
needed in these regions.  Therefore, we do not want to spend too much
work on $\delta\Omega$. Using the modified boundary functions, we
extrapolate ($d$-linearly) towards $\delta\Omega$, see the discussion
of the modified basis functions above.

Second, the modifications are especially well-suited if the function
to interpolate can be separated into a sum of one-dimensional
functions. Assume that the
likelihood $\LL$ was a perfect product of one-dimensional Gaussians,
\begin{equation}
  \LL(\vx) = c \cdot \e^{-a_1 (x_1-\mu_1)^2 - \ldots -a_d (x_d-\mu_d)^2}\,,
\end{equation}
centred at $(\mu_1,\ldots,\mu_d)^T$.
Then the interpolation of the log-likelihood $\LLL$ reduces to $d$
one-dimensional problems,
\begin{equation}
  \LLL(\vx) = \ln c + \sum_{k=1}^d f_k(x_k)\,,
\end{equation}
with
\be
f_k(x_k) = -a_k (x_k-\mu_k)^2\,,
\ee
separating into a constant term plus a sum of functions that are
constant in all directions but one.

Keeping in mind that the one-dimensional basis function on level 1,
$\varphi_{1,1}(x)$, is constant (\cf
Figure \ref{fig:basisfunktionen_ausgeklappt}), this simplifies the
interpolation task. The $d$-dimensional basis function on level
$\vone$, $\varphi_{\vone,\vone}(\vx)$, serves as an offset. (Only if
$(\mu_1,\ldots,\mu_d)^T$ is the centre of $\Omega$,
$\alpha_{\vone,\vone}\varphi_{\vone,\vone}(\vx)$ exactly expresses
$\ln c$.) Additionally, it is sufficient to spend only grid points on
the main axes of the sparse grid (level 1 in all dimensions but one)
to approximate the remaining one-dimensional contributions $f_k(x_k)$
arbitrarily well:
\begin{eqnarray} \nonumber
  u(\vx) &=&
  \underbrace{\alpha_{\vone,\vone}\varphi_{\vone,\vone}(\vx)}_{\normalsize
    \ln c}
  \\
&&+ \sum_{k=1}^d \underbrace{\left(\sum_{l_k,i_k}
    \alpha_{\vl,\vi} 
  ~\varphi_{l_k,i_k}(x_k) \!\!\!\!\! \prod_{1\leq j\leq d, j\neq k} \!\!\!\!\!
  \varphi_{1,1}(x_j)\right)}_{\normalsize f_k(x_k)}.
\end{eqnarray}

Of course, $\LL$ is not a perfect product of one-dimensional
Gaussians; grid points that do not lie on the sparse grid's main axes
account for the additional mixed (correlated) terms of $\LLL$. Given
that in sparse grids a large amount of points lie on the main
axes, this mechanism works very well---the better, the less
correlation between the different parameters exists. 

In order to take as much advantage as possible of the effects
described above, we introduce a parameter transformation in the
following section, for which the new parameters are less correlated.
However, the fact that the interpolation using the standard
parameters---which have much stronger correlations---works as well,
spending just more grid points, will show that the sparse grid
approach does not depend on this argumentation: Sparse grids can make
use of such properties but do not rely on them.

\subsection{Normal parameters}\label{sec:trafo}

The set of cosmological parameters describing the
$\Lambda$CDM model consists of the Hubble constant, $h \equiv
\frac{H_0}{100\,\km/(s \Mpc)}$, the density parameter of vacuum energy,
$\Omega_\Lambda$, 
the ones of baryons, $\Omega_b$, and of matter (baryonic + dark),
$\Omega_m$, the optical depth to the last scattering surface, $\tau$,
the scalar spectral index of the primordial power spectrum, $n_s$, and
the scalar initial amplitude, $A_s$.  We will refer to these parameters
as cosmological parameters. For a more detailed description of the
cosmological parameters, we refer to
\cite{coles_lucchin}.  In the literature, there have been several
attempts to transforming these parameters into a set of parameters
that mirror the various physical effects on the CMB power spectrum
\citep{hu,kosowsky}. In \cite{normal_params}, a set of
parameters is provided in which the likelihood-surface of the CMB is
well approximated by a multivariate Gaussian with the major axes
aligned with the coordinate axes. In this work, we use the
parameters given by \cite{cmbfit}, where the parameter set of
\cite{normal_params} is combined with the other parameter sets
mentioned, in order to bring the major axes of the likelihood surface
even closer to the coordinate axes. The new parameters are then
$\{\Theta_s, h_2, h_3, t, A_*, Z\}$, which we refer to as normal
parameters. When working with the latter, the logarithm of the
likelihood is well-approximated by a sum of one-dimensional parabolas
in the different parameters. The basis functions introduced above are
therefore ideally adapted to this problem.  In the following, we
repeat the definitions of the normal parameters for convenience.

The first parameter of our set is the angle subtended by the acoustic scale
\be\label{thetaS}
\Theta_s \equiv \frac{r_s(a_\Ls)}{D_A(a_\Ls)}\frac{180}{\pi}\,,
\ee
where the index $\Ls$ denotes the time of last scattering,
$D_A(a_\Ls)$  stands for the comoving angular diameter
distance to the surface of last scattering (which we will come back to later),
and $r_s(a_\Ls)$ is the
comoving sound horizon at last scattering,
\be
r_s(a_\Ls) \equiv \int_0^{t_\Ls} \frac{c_s(t)}{a(t)}dt\,.
\ee
Here, $c_s(t)$ denotes the sound speed for the baryon-photon-fluid at time $t$,
which is well approximated by
\be
c_s(t)^2 \approx \frac{1}{3}(1+3\frac{\rho_b}{\rho_\gamma})^{-1}\,,
\ee
with the index $b$ standing for baryons and the index $\gamma$ for
photons.
Using the Friedmann equations and ignoring the vacuum energy at
last scattering, $r_s(a_\Ls)$ can be shown to be \citep{cmbfit,kosowsky}
\be
r_s(a_\Ls) =
\frac{2\sqrt{3}}{3H_0\sqrt{\Omega_m}}\sqrt{\frac{a_\Ls}{R_\Ls}}\ln\frac{\sqrt{1+R_\Ls}+\sqrt{R_\Ls+r_\Ls
    R_\Ls}}{1+\sqrt{r_\Ls R_\Ls}}\,,
\ee
where
\bea
R_\Ls &\equiv& \frac{3\rho_b(a_\Ls)}{4\rho_\gamma(a_\Ls)} = 30\, w_b
\left(\frac{z_\Ls}{10^3}\right)^{-1}\,,\\ 
r_\Ls &\equiv& \frac{\rho_r(a_\Ls)}{\rho_m(A_*)} =
0.042 \,w_m^{-1}\left(\frac{z_\Ls}{10^3}\right)\,.
\eea
The index $r$ stands for radiation, i.e., $\,\rho_r$ consists of the sum
of photon and neutrino energy densities, and the index $m$ is used for matter
(baryons + dark matter). We define $w_m \equiv \Omega_m h^2$ in
the same way as $w_b \equiv \Omega_b h^2$. 
The redshift at last scattering, $z_\Ls$, is well approximated by \citep{hu}
\bea
z_\Ls &=& 1048\,(1+0.00124 \, w_b^{-0.738})(1+g_1w_m^{g_2})\,,\\
g_1 &\equiv& 0.0783 \, w_b^{-0.238}(1+39.5 \, w_b^{0.763})^{-1}\,,\\
g_2 &\equiv& 0.560\,(1+21.1w_b^{1.81})^{-1}\,.
\eea
As already mentioned, $D_A(a_\Ls)$ in (\ref{thetaS}) denotes the
comoving angular diameter 
distance to the surface of last scattering and is given by
\be
D_A(a_\Ls) = \frac{c}{H_0}\int_{a_\Ls}^1\frac{1}{\sqrt{\Omega_\Lambda
    \tilde a^4+\Omega_m\, \tilde a+\Omega_r}}d\tilde a\,. 
\ee

The second and third parameters in our set are the ratios of the second and the
third peak to the first peak in the $C_l^{T}$ spectrum of the CMB \citep{hu},
where the tilt-dependence is factored out \citep{page_params},
\bea \nonumber
h_2 &\equiv&  0.0264 \,w_b^{-0.762}\\
&&\exp\left(-0.476\,[\ln(25.5\, w_b+1.84 \,w_m)]^2\right)\,,\\ \nonumber
h_3 &\equiv& 2.17\left(1+\left(\frac{w_b}{0.044}\right)^2\right)^{-1}\\ 
&& w_m^{0.59}  \left(1+1.63\left
(1-\frac{w_b}{0.071}\right)w_m\right)^{-1}\,. \label{h3} 
\eea
We use the tilt parameter given by \cite{cmbfit}, which is a slightly modified
version of the one in
\cite{normal_params} in order to minimise the correlation with $w_b$:
\be
t \equiv \left(\frac{w_b}{0.024}\right)^{-0.5233}2^{n_s-1}\,.
\ee
The amplitude parameter is 
\be
A_* \equiv \frac{\tilde A_s}{2.95\times
  10^{-9}}{\mathrm e}^{-2\tau}\left(\frac{k}{k_p}\right)^{n_s-1}w_m^{-0.568}\,,
\ee
where $k_p = 0.05\mathrm{Mpc}^{-1}$ denotes the pivot point.
The normalisation factor of $2.95\times 10^{-9}$ comes in because we use the
scalar amplitude $\tilde A_s$ of {\small CMBEASY}, which is defined as the
primordial power of the curvature fluctuations evaluated at the pivot point,
$\tilde A_s \equiv \Delta_R^2(k_p)$. It is related to the scalar amplitude
$A_s$ of {\small CMBFAST}, which is used in \cite{cmbfit}, by $A_s =
\frac{\tilde A_s}{2.95 \times 10^{-9}}$ \citep[\cf][]{verde}. 
Finally, we use the physical damping due to the optical depth to last
scattering as our last parameter: 
\be
Z \equiv {\mathrm e}^{-\tau}\,.
\ee

In order to construct the interpolation of the likelihood surface, we need the
transformation that maps the normal parameters back onto 
cosmological parameters. The reason for this is the way we construct the
interpolation: 
Our sparse grid algorithm chooses the normal parameters where it wants to
refine the grid, which we then need to transform into cosmological
parameters to run {\small CMBEASY} and the WMAP-likelihood code. 
Our technique of inverting the parameter transformation is presented in
appendix \ref{app:sparse}.

\subsection{Generation of test set and choice of interpolation range}
\label{sec:test_set}

For choosing the parameter range in which to construct the
interpolation, we have run MCMCs containing about 50,000 points at a
temperature of $T=3$. That is, in the Metropolis
algorithm we choose the transition probability $a(\vx,\vy)$ from a point
$\vx$ in the chain to a new point $\vy$ to be 
$ a(\vx,\vy) \equiv \min \left\{\left(\LL(\vy) /
\LL(\vx)\right)^{\frac{1}{T}}, 1 \right\} $. Using this
transition probability with $T=1$ results in the usual Metropolis
algorithm, whereas choosing $T=3$
allows us to explore a larger parameter range than with
the regular algorithm. These chains covered a region reaching out to
about 25 log-likelihoods around the peak. 

The optical depth to the
last scattering surface, $\tau$, which can be determined from the CMB
polarization, is not well-constrained by the WMAP polarization
data due to their low signal-to-noise ratio. Therefore, when running
the MCMCs at 
$T=3$, we had to restrict $\tau$ to the physically meaningful range
$\tau \geq 0$. This restriction corresponds to $Z \leq 1$ for the normal
parameters. In the case of the normal parameters in 7 dimensions, we had to
additionally restrict the intervals to $h_2 \leq 0.52$ and $h_3 \geq
0.38$, which is the range in which the
parameter transformation is invertible.
Furthermore, we chose to restrict our set of points
to be within the 25 log-likelihood region around the peak.

In order to roughly centre our intervals at the maximum of the
log-likelihood function, we have determined the latter using a few
runs of a simple simplex search.\footnote{We were running several
  simplex searches and chose the result with the highest value of the
  log-likelihood. The runs did not all converge to exactly the same
  point, which we think was due to numerical issues (the
  log-likelihood was presumably not completely convex, which could be
  due to the dips we will mention in Sec.\ \ref{sec:adaptive_grids}).}
The interval boundaries were then defined as the box centred at the
maximum which contains all points of the above-described chains.
Note that it is not important for the accuracy of the
interpolation that the intervals are well-centred at the maximum.
Note further that we have used this set of points as a
test set for comparing our interpolation with the real log-likelihood.

\subsection{Results}\label{sec:results}

\begin{figure}
\centering
\includegraphics[angle=270,width=0.4\textwidth]{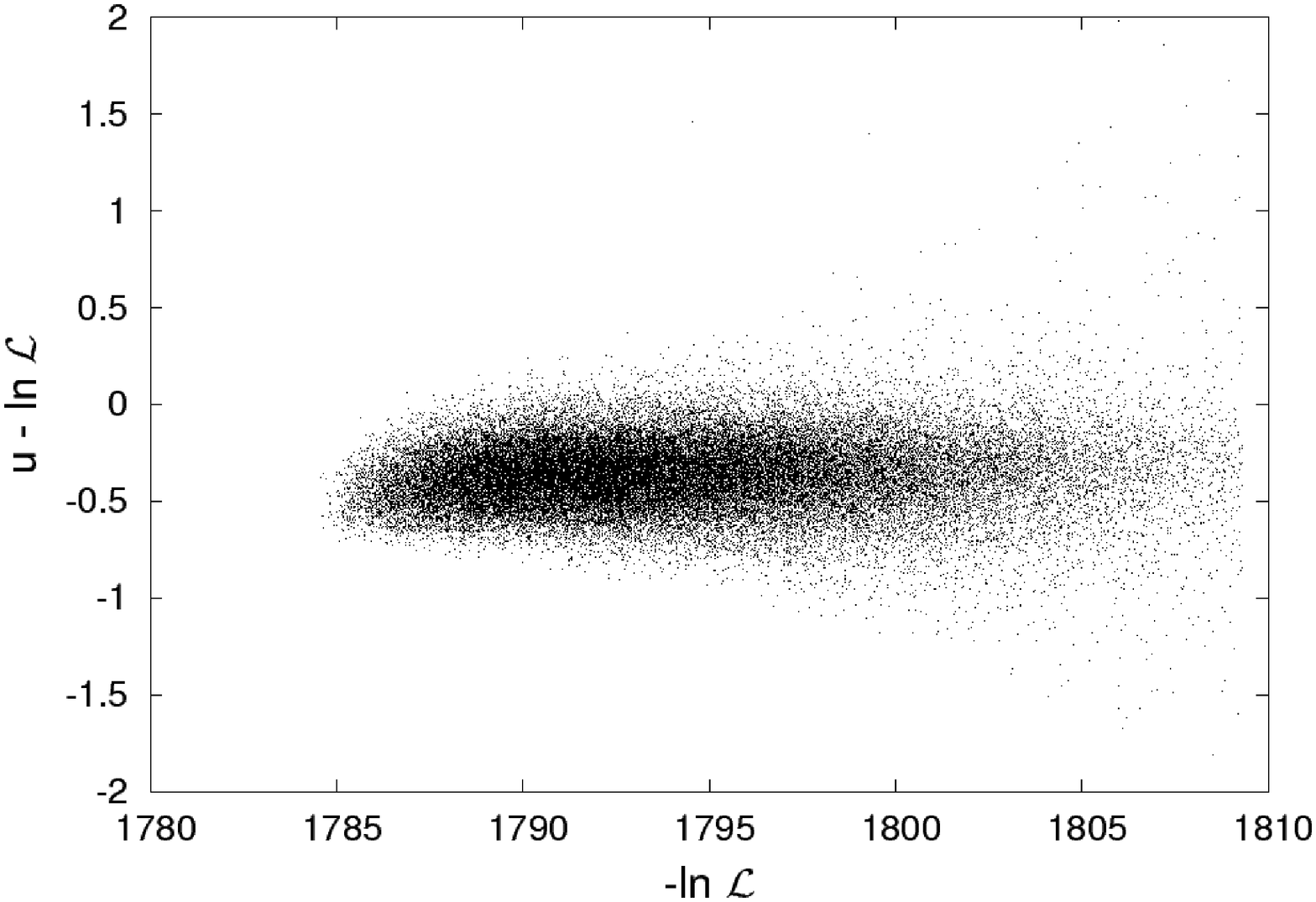}
\includegraphics[angle=270,width=0.4\textwidth]{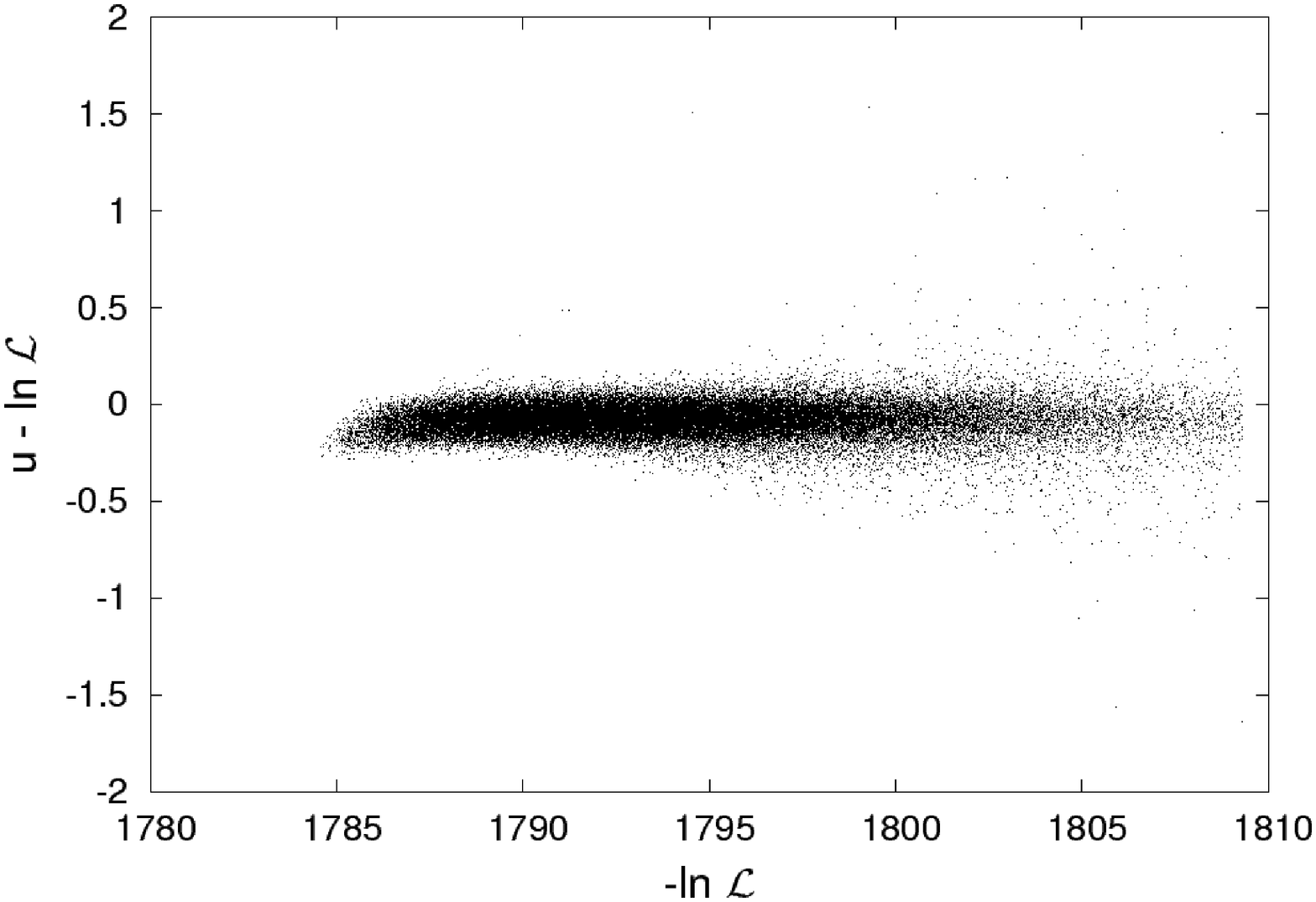}
\caption[Scatterplots, 6 dimensions, normal parameters]{Absolute error
  of the interpolation with respect to the real 
  log-likelihood in 6 dimensions for an interpolation with a sparse
  grid of level 5 (top panel) and of level 6 (bottom
  panel) for normal parameters. 
} 
\label{fig:dirk_newbounds_scatter_6dim_l5_T3}
\end{figure}
\begin{figure}
\centering
\includegraphics[angle=270,width=0.4\textwidth]{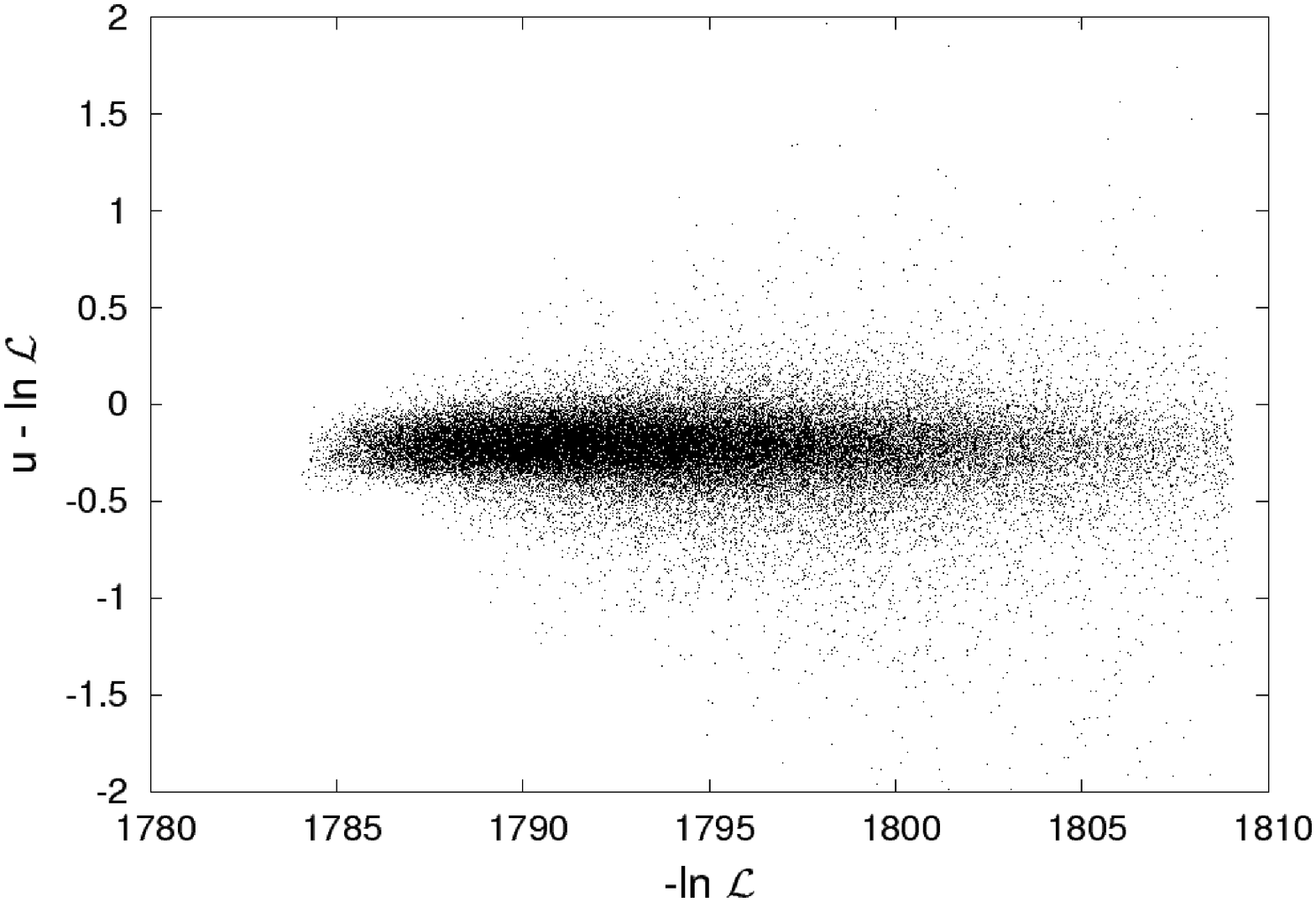}
\includegraphics[angle=270,width=0.4\textwidth]{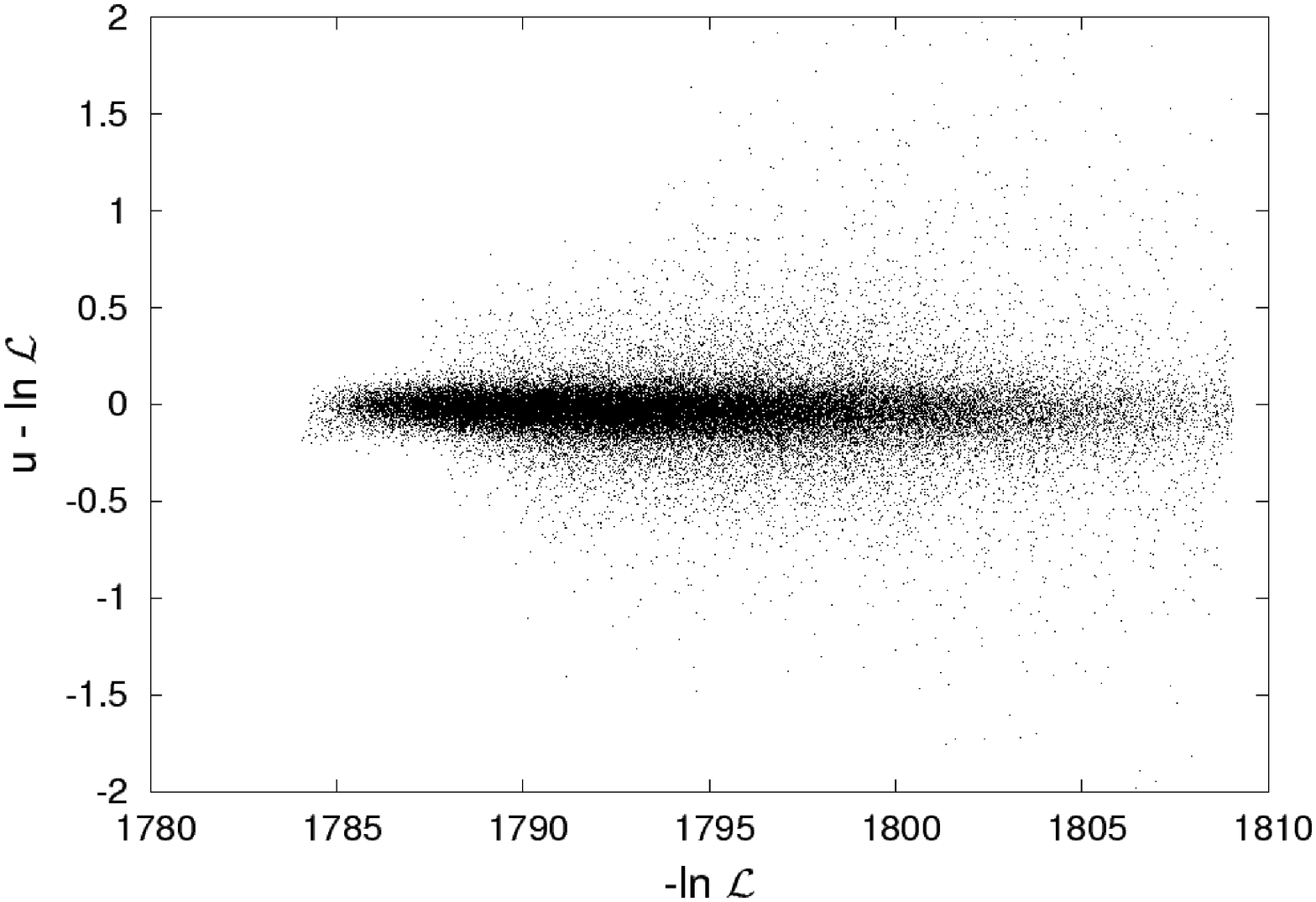}
\caption[Scatterplots, 7 dimensions, normal parameters]{Absolute error
  of the interpolation with respect to the real 
  log-likelihood in 7 dimensions for an interpolation with a sparse
  grid of level 6 (top panel) and of level 7 (bottom panel) for normal
  parameters.} 
\label{fig:salsa_scatter_7dim}
\end{figure}

We have interpolated the log-likelihood of the WMAP 5 year data in the
6-dimensional normal parameter space described in Sec.\ \ref{sec:trafo}. 
The same has been done for a 7-dimensional model, in which we
have chosen the running of the spectral index of the primordial power
spectrum, $\alpha$, as an additional parameter. Constructing the
interpolation can be parallelised to an arbitrary degree, according 
to the available computational resources. 

For the 6-dimensional model, we plot the absolute error of the
log-likelihood, $(u - \LLL)$, against the negative WMAP
log-likelihood, $(-\ln \LL)$, for the points in the test set in
Figure \ref{fig:dirk_newbounds_scatter_6dim_l5_T3}. We have used an
interpolation on a sparse grid of level $n=5$ (consisting of 2561
grid points) in the top panel, and of level $n=6$ (consisting of
10625 grid points) in the bottom panel. One clearly sees the
improvement in accuracy when increasing the grid level from $n=5$ to
$n=6$.  Figure \ref{fig:salsa_scatter_7dim} shows the same plot for the
7-dimensional model, for grid level $n=6$ (18943 grid points) in the
top panel and $n=7$ (78079 grid points) in the bottom panel. We again
see the improvement in accuracy with increasing refinement
level. However, the additional parameter $\alpha$ is quite strongly
correlated with many of the other parameters, whereas the correlations
between the normal parameters in 6 dimensions are reduced to a
minimum.  We therefore have to increase the grid level by one in 7
dimensions, in order to obtain results comparable to the 6-dimensional
ones. In both figures, we note a systematic negative offset of the
interpolation with respect to the real function, which becomes less
severe for the higher refinement levels. This offset is due to the
fact that we construct a $d$-linear interpolant of a convex function,
which systematically lies below the function. This could be easily
coped with by adding a small offset to $\alpha_{\vone,\vone}$ after
the interpolation, or, even better, by using piecewise polynomial
instead of the piecewise linear basis functions. We leave the usage of
piecewise polynomial basis functions, which promise to be well-adapted
to the log-likelihood, for future work.

Note that we have restricted the plot range to [-2,2], because only
0.1 per cent or less of the points lie outside this range.\footnote{In 6
  dimensions, the number of points outside this range is 0.02 per cent
  (0.003 per cent) for $n=5$ ($n=6$); in 7 dimensions, it is about 0.1 per cent for
  both grid levels.}  Almost all of these points lie in the corners of
$\Omega$ due to relatively strongly correlated parameters. These are
the regions in parameter space where the 25 log-likelihood range
around the peak extends to the interval boundaries. Due to the
extrapolation we use close to the boundaries (\cf the end of
Sec.\ \ref{sec:sg_basics}), we obtain relatively large uncertainties
in those regions, which do not affect the one-dimensional projections
of the likelihood function, though. The uncertainties can be further
reduced, spending (adaptively) more grid points in those
regions, see also the discussion about adaptivity in
Sec.\ \ref{sec:adaptive_grids}. 

For the 6-dimensional interpolation with a sparse grid of level 6,
2.5 per cent of the test points have an absolute error $>0.25$ in the
log-likelihood, and 0.03 per cent of the test points have an absolute
error $>1$. In 7 dimensions and for refinement level 7, the
corresponding numbers are 9 per cent and 0.5 per cent,
respectively. This is a higher level of 
accuracy as reached by Pico \citep{pico}, for which about 90 per cent
of the points in a region reaching out to 25 log-likelihoods around
the peak have been calculated with an absolute error below
0.25. However, we note that these numbers for Pico are valid for a
9-dimensional parameter space, whereas we work in 6- and 7-dimensional
spaces and leave the extension to higher-dimensional models to future
work. But we also note that in all settings where a systematic offset
in the interpolation error can be observed, it is sufficient to reduce
the offset to improve our results significantly, in particular for
interpolations on lower levels (see, e.g., the scatterplot for the
6-dimensional model and grid level 5,
Figure \ref{fig:dirk_newbounds_scatter_6dim_l5_T3}).

\begin{figure}
\centering
\includegraphics[width=0.2\textwidth]{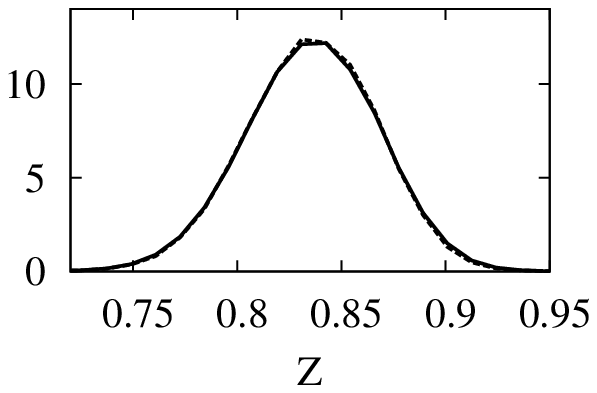}
\includegraphics[width=0.2\textwidth]{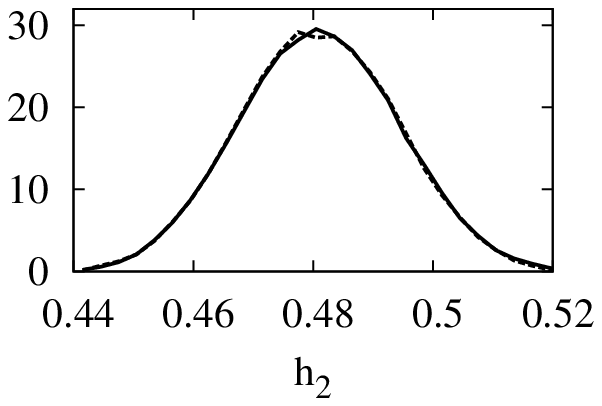}
 \includegraphics[width=0.2\textwidth]{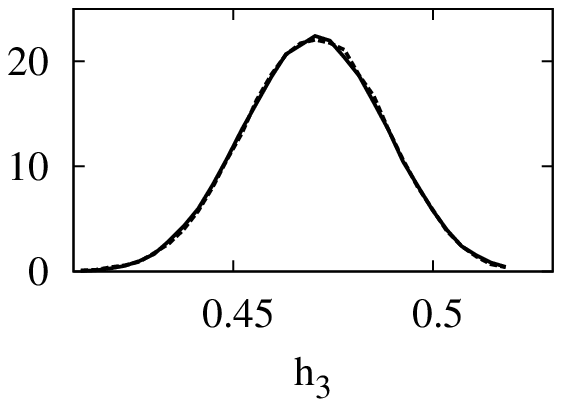}
\includegraphics[width=0.2\textwidth]{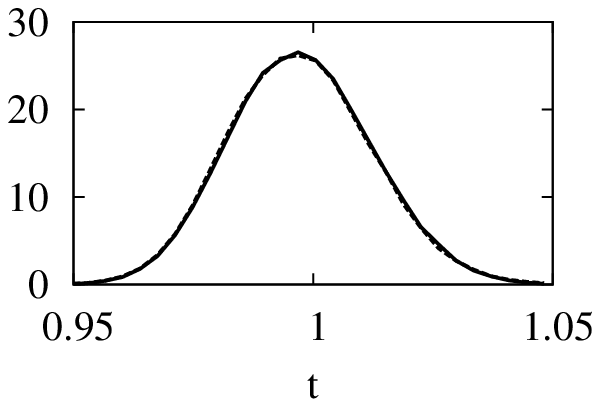}
\includegraphics[width=0.2\textwidth]{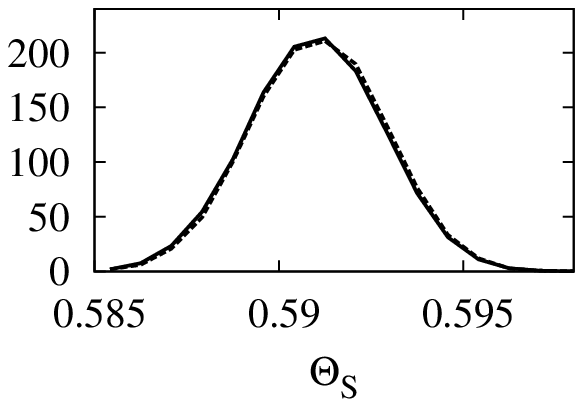}
\includegraphics[width=0.2\textwidth]{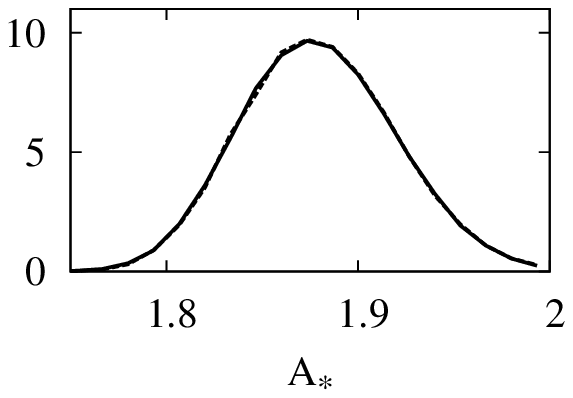}
\includegraphics[width=0.2\textwidth]{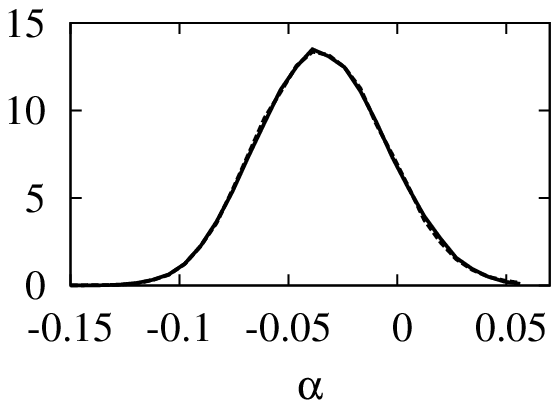}
\caption[Histograms, 7 dimensions, normal parameters]{Comparison of
  the one-dimensional projections of the 7-dimensional 
  WMAP 5 year likelihood (solid) and its interpolation (dashed) using
  a sparse grid of level $n=6$ (consisting of 18943 grid points)
  for normal parameters. The curves match almost perfectly.}
\label{fig:salsa_7dim_level6}
\end{figure}

We have projected both the interpolation and the WMAP likelihood
function using MCMCs of about 150,000 points,
and compare the results for the 7-dimensional model for grid level
$n=6$ in Figure \ref{fig:salsa_7dim_level6}. We reproduce the 
projected one-dimensional likelihood curves almost perfectly. The
results for the 6-dimensional model for grid level $n=5$ are very
similar to the 7-dimensional ones. Note that we obtain these excellent
results for the 6-dimensional model using only 2561 grid points.
The visual comparison of our results with the projected one-dimensional
likelihoods obtained by CosmoNet \citep{cosmonet_1,cosmonet} shows
that we reproduce the original curves with a higher
accuracy than the latter. Note also that our interpolation is
constructed in a rather wide region, encompassing about 25
log-likelihoods around the peak, whereas in \cite{cosmonet} the region
in which $\LLL$ was fitted covers only $4\sigma$ around the peak for
the combined likelihood of CMB and LSS. This corresponds to a region
of about 8 log-likelihoods around the peak for the combined
likelihood, and even less when using only the CMB likelihood.

\begin{figure}
\centering
\includegraphics[angle=270,width=0.4\textwidth]{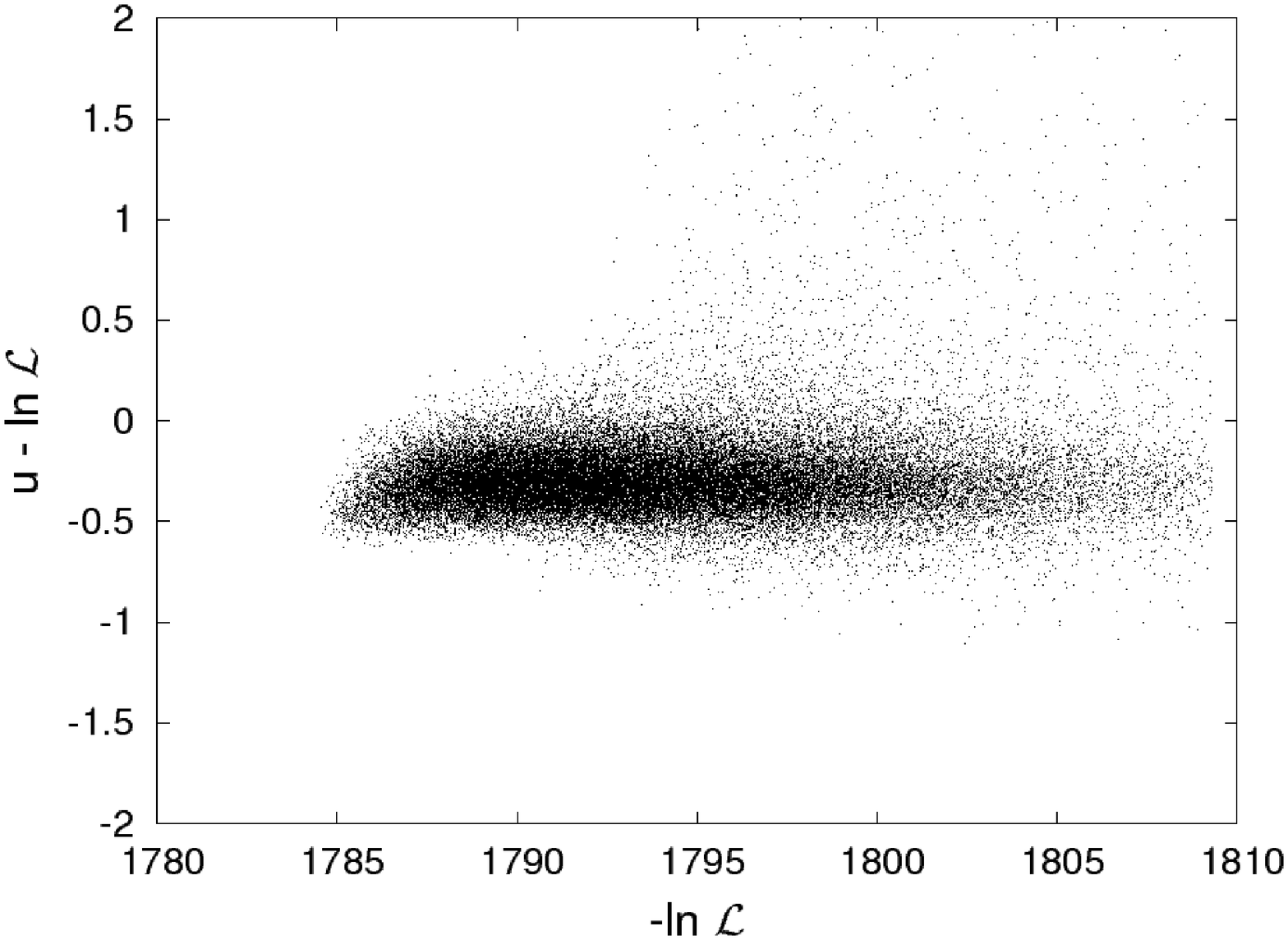}
\includegraphics[angle=270,width=0.4\textwidth]{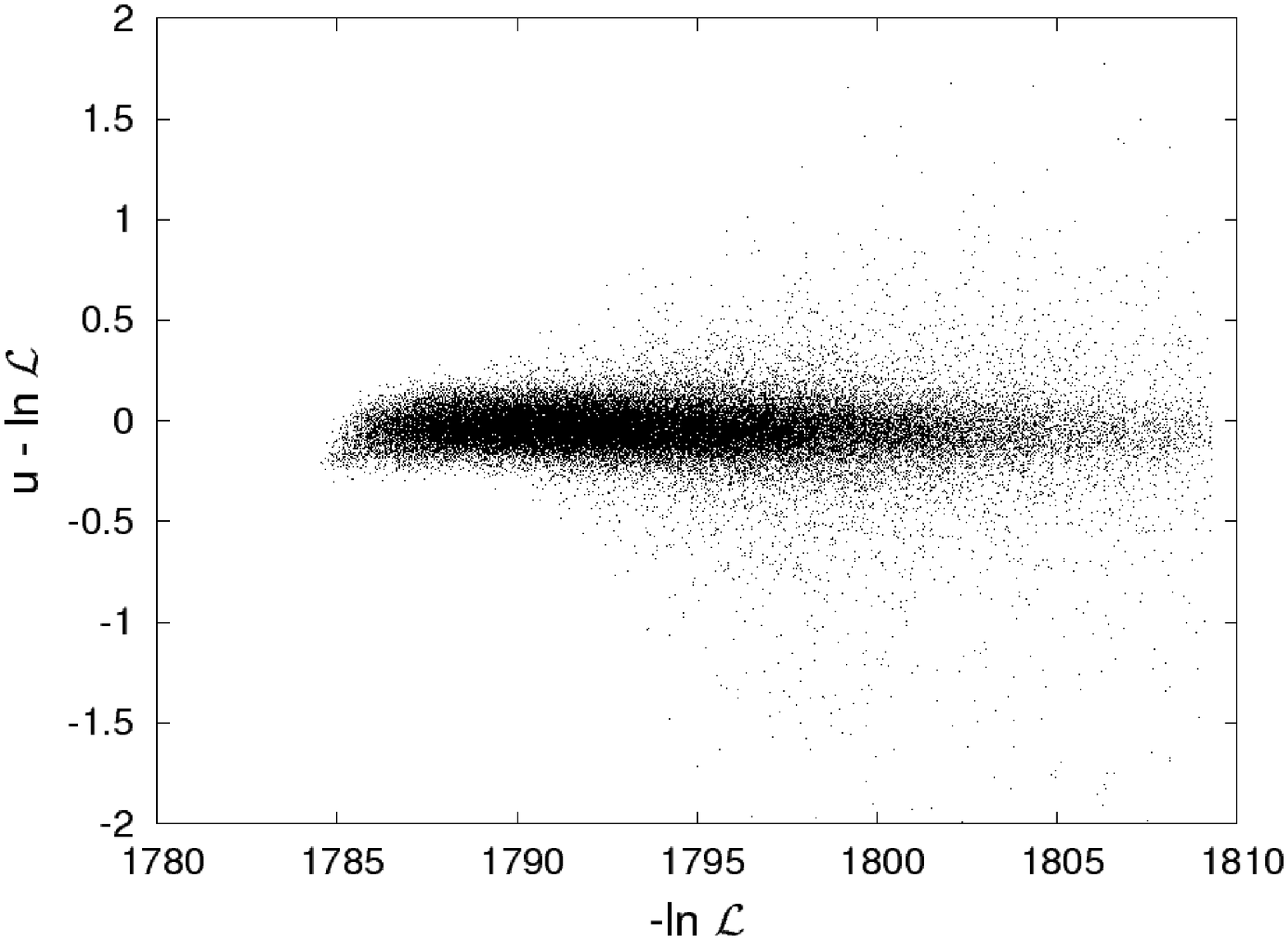}
\caption[Scatterplots, 6 dimensions, standard parameters]{Absolute
  error of the interpolation with respect to the real 
  log-likelihood in 6 dimensions for an interpolation with a sparse
  grid of level 6 (top panel) and of level 7 (bottom
  panel) for standard parameters. 
} 
\label{fig:dirk_cosmos_scatter_6dim_l5_T3}
\end{figure}

\begin{figure}
\centering
\includegraphics[angle=270,width=0.4\textwidth]{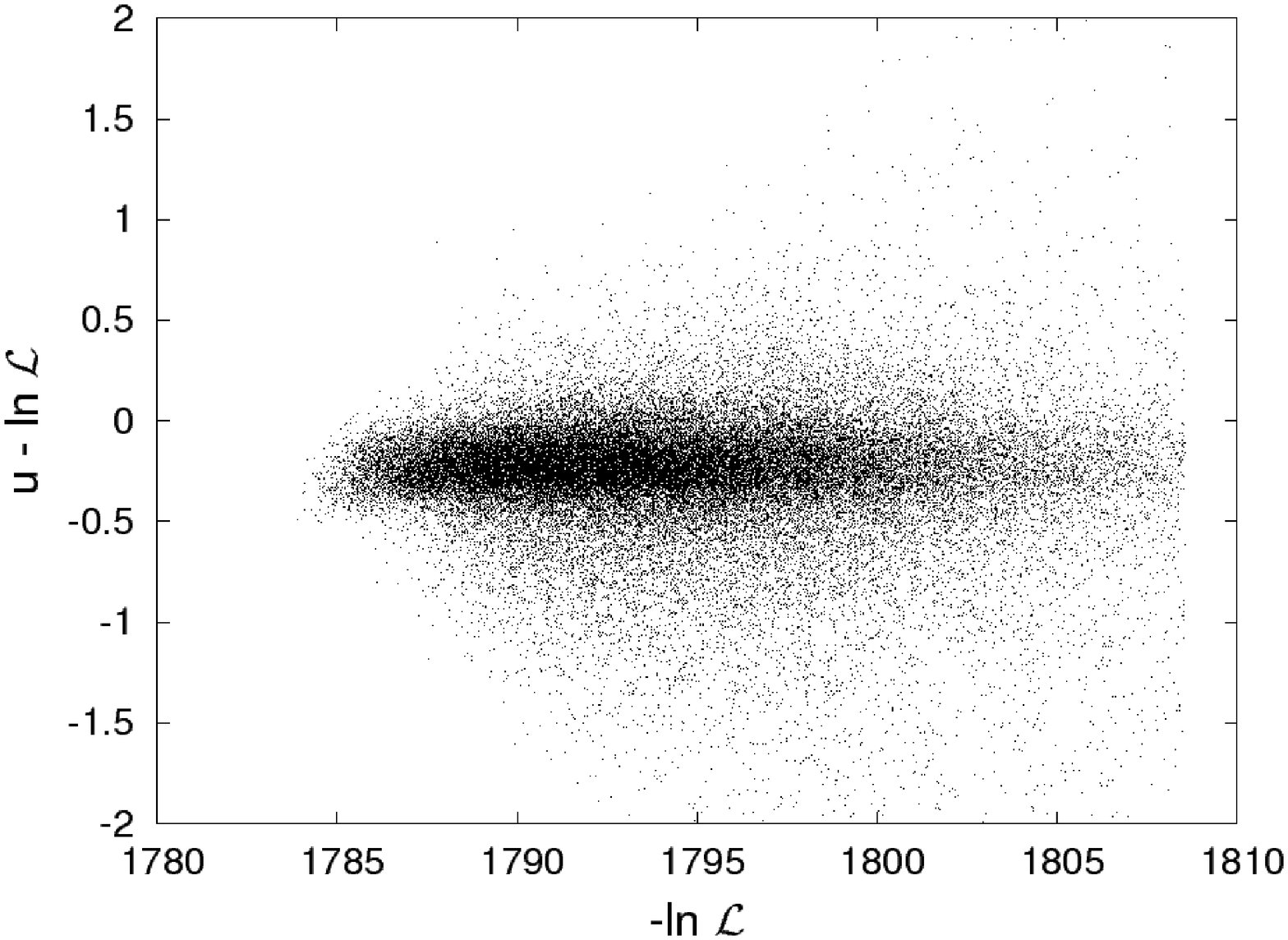}
\includegraphics[angle=270,width=0.4\textwidth]{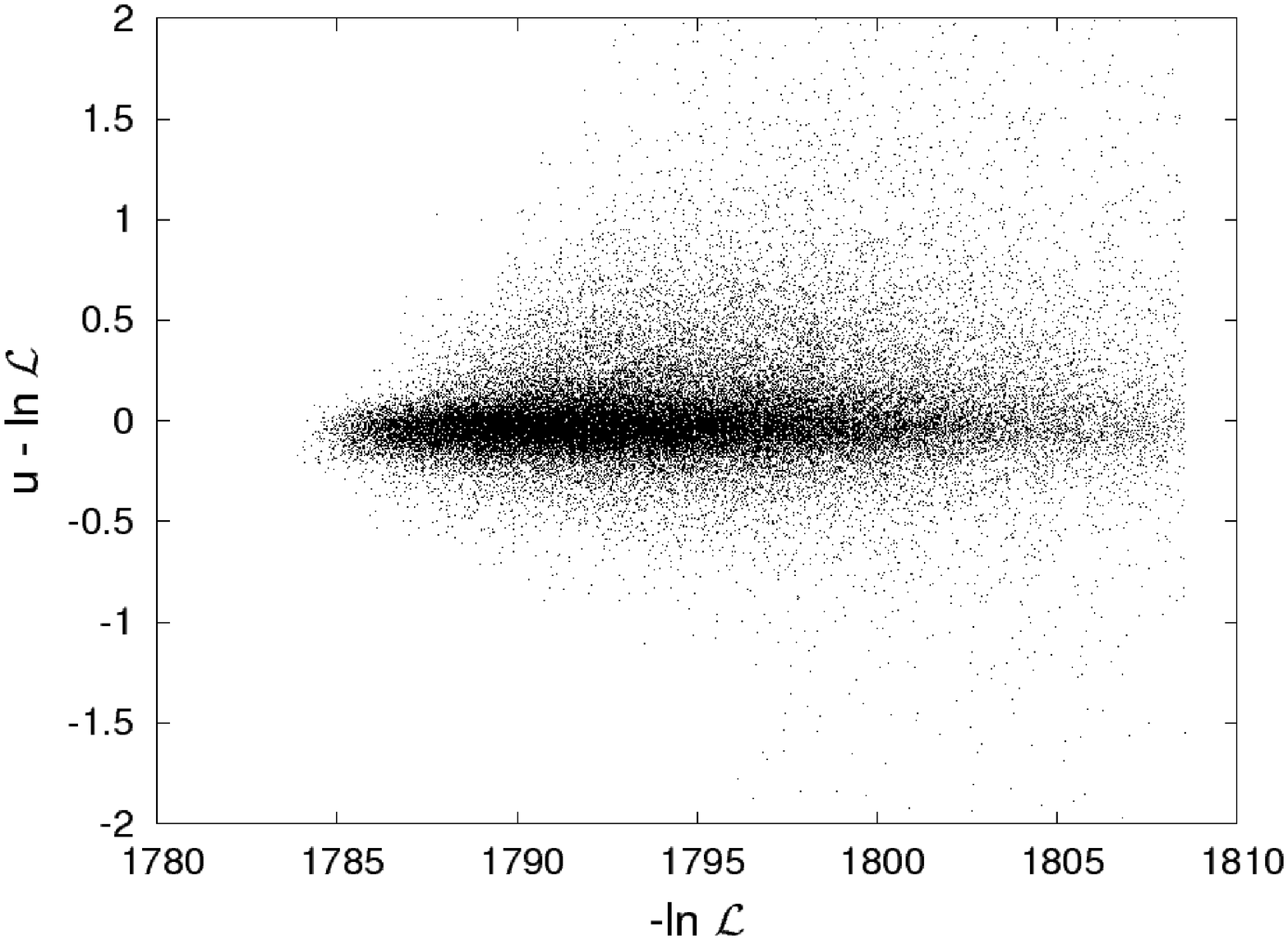}
\caption[Scatterplots, 7 dimensions, standard parameters]{Absolute
  error of the interpolation with respect to the real 
  log-likelihood in 7 dimensions for an interpolation with a sparse
  grid of level 7 (top panel) and of level 8 (bottom
  panel) for standard parameters. 
} 
\label{fig:dirk_cosmos_scatter_7dim_l5_T3}
\end{figure}

Consider now the 
interpolation of the WMAP likelihood surface using directly the
standard parameters, which are used by default when doing
cosmological parameter sampling with the MCMC driver from {\small
  CMBEASY} \citep{analyzethis}: $\{w_m,w_b,h,\tau,n_s,\ln(10^{10}A_s)
- 2\tau\}$, to which we again add $\alpha$ as an additional parameter
in the 7-dimensional case. Working with these parameters has the
advantage that we do not have to restrict 
ourselves to the parameter range in which the parameter transformation
is invertible. However, the problem is now less adapted to our choice
of basis functions, due to the stronger correlations between the different
parameters. We therefore pay the price of having to increase the grid level
by one in this case in order to reach an accuracy as good as before. We show the
absolute error of the log-likelihood, $(u - \LLL)$, 
against the negative WMAP log-likelihood, $(-\LLL)$, for the
6-dimensional model for grid level $n=6$ (10625 points) and $n=7$
(40193 points)
in Figure \ref{fig:dirk_cosmos_scatter_6dim_l5_T3},\footnote{Here,
  about 0.3 per cent (0.2 per cent) of the points in the test set lie
  outside the chosen plot-range for the grid of level $n=6$ ($n=7$).}
and for the 7-dimensional one for $n=7$ (87079 points) and $n=8$
(297727 points) in
Figure \ref{fig:dirk_cosmos_scatter_7dim_l5_T3}.\footnote{About 
  1 per cent of the points in the test set lie outside the
  chosen plot-range for the grid of both level $n=7$ and level $n=8$.}
For the 6-dimensional (7-dimensional) interpolation with a sparse grid
of level $n=7$ ($n=8$), the fraction of test points with absolute
error $>0.25$ in the log-likelihood is 6 per cent (20 per cent),
and 0.5 per cent (2.5 per cent) for an absolute error $>1$. 
Again, the one-dimensional projections of the 6-dimensional case for 
level $n=6$ and of the 7-dimensional case for level $n=7$ are very
similar to the ones in Figure \ref{fig:salsa_7dim_level6}, and
are thus not shown. 


We have tested the evaluation time of our interpolation by evaluating
a sparse grid interpolant of level 6 in 6 dimensions for 2,000,000
points randomly chosen from within $\Omega$.
On a conventional desktop computer (Intel chipset, 2.8\,GHz), this took
about $92\,\mu s$ per point, including the random generation of the
point. In 7 dimensions on the same level we have twice as many grid
points and one dimension more, which doubles the evaluation time to
$189\,\mu s$. 
For CosmoNet and Pico, the 
evaluation of a 6-dimensional model is specified to take about
$10\, \mu s$ and $250\, \mu s$, respectively
\citep{cosmonet}. Note that we do not know on which hardware
the evaluation times of CosmoNet and Pico have been measured, which
makes a comparison hardly possible. Note further that our code to
evaluate a sparse grid function is not optimised for fast evaluation
times and that there is still room for improvement. In any case, for
all of these codes the bottleneck in cosmological parameter sampling
is now the MCMC algorithm itself rather than the evaluation of the
likelihood, at least with the MCMC driver used in this work
\citep{analyzethis}.

Note that in 7 dimensions, we need significantly more grid points than
in 6 dimensions, since the additional parameter $\alpha$ strongly
correlates with the other parameters, and we thus need to increase the
grid level by one to obtain good results. As the storage requirements
are rather low, this mainly increases the number of evaluations that
are needed for constructing the sparse grid interpolation. As already
stated before, though, the construction of the interpolation can be
parallelised to an arbitrary degree, according to available computational
resources, so that this should not be an issue. To store the
interpolant for a regular sparse grid in $d$ dimensions for level $l$
with $N$ grid points, we would only need $N$ doubles for the
coefficients and two integers to remember both $d$ and $l$, leading to
$(N+1)\cdot8$ Bytes. For
adaptively refined sparse grids, we additionally have to store at
least which grid points have been refined, requiring slightly more
storage. For current hardware architectures, the size of the memory is
therefore not a limiting factor for our application.

We have shown in this section, that the interpolation of the
log-likelihood with regular sparse grids provides excellent and
competitive results in both 6 and 7 dimensions, even when
interpolating in the rather wide range covering about 25 log-likelihoods
around the peak. 
The memory requirements to store the sparse grid are
rather low and the evaluation times are very fast. We have
demonstrated that the one-dimensional projections of the
interpolants match the original ones almost perfectly, and that the
scatterplots of the errors exhibit only a small fraction of grid
points with higher errors. For standard parameters, more grid points
have to be spent to obtain the same accuracy than for normal
parameters, as the latter ones correlate less. In the next section,
we will therefore focus on an adaptive extension of the method used
so far to further reduce the number of grid points that are
required.

\subsection{Improvements with adaptive sparse grids}\label{sec:adaptive_grids}

As it has already been mentioned, the log-likelihood is not a perfect
sum of one-dimensional functions. The different parameters contribute
differently to $\LL$ and correlate more or less with each other. It is
therefore reasonable, especially when using the standard parameters
which correlate more, to employ adaptivity, spending more grid points
in critical regions and less grid points elsewhere. 
In this section, we demonstrate the utility of adaptivity by
showing some first results as a proof of concept. As they can clearly
be improved further, we leave a thorough study of adaptive sparse
grids for the interpolation of $\LLL$ to future work. We start
by specifying how to refine, we formally derive a suitable criterion
that can be used to specify where to refine as well as to measure
the quality of an interpolation, and we finally provide results that
show how adaptivity can improve the interpolation obtained for
regular sparse grids.

Employing adaptivity, one can attempt to either obtain better results
fixing roughly the number of grid points used, or to achieve a similar
accuracy using less grid points. In the following, we show results for
the former, tackling the 7-dimensional example using the standard
parameters on level 7 with 78079 grid points presented above. We start
with a regular sparse grid of some low level and refine grid points,
creating all $2d$ children in the hierarchical structure (if possible)
each, until the grid size exceeds 78000 grid points. In settings where
the contributions of the dimensionalities differ significantly, it can
be useful to start with level 2 to allow dimensional adaptivity,
neglecting unimportant dimensions; here, the grid points on low levels
will be created in any case, so we can start with a sparse grid on
level 5, e.g., to save on the number of adaptive steps.

Choosing a suitable refinement criterion, it can be determined whether
to refine in a broad way (close to regular sparse grids) or in a more
greedy way in the sparse grid's hierarchical tree structure. It is
reasonable to take the surpluses of the grid points into account as
they contain the local information about the functions, i.e., if the
function has a high gradient locally. Furthermore, they decay quickly
with increasing level-sum in the convergent phase.
The mere surplus-based criterion, refining the grid
points with the highest absolute value of the surplus first, is known
to tend to minimise the $L_2$-norm of the error. As we do not spend
grid points on the domain's boundary, and as we are extrapolating
towards the boundary, the biggest surpluses per level can be found
for the modified basis functions which are adjacent to the boundary. A
mere surplus-based criterion will therefore only refine towards the
boundary. This reduces the error especially for sampling points with a
high error in the scatterplots, as they are located towards the
boundary of the domain.

In the following, we theoretically derive a refinement criterion which
is better suited to our problem than the purely surplus-based one.  In
order to maximise the information our interpolation contains about the
real likelihood, we attempt to minimise the Kullback-Leibler distance
$d_{KL}$ between the interpolation and the likelihood function,
\begin{eqnarray} \nonumber
d_{KL} &\equiv& \int d^dx\, \LL(\vx) \ln \frac{\LL(\vx)}{\exp(u(\vx))} \\
&=& \int d^dx\, \LL(\vx) \left(\LLL(\vx) - u(\vx)\right)\,,
\label{KL_orig}
\end{eqnarray}
which is defined for two normalised probability distributions $\LL$
and $\exp(u)$.  Let us now derive the refinement criterion we obtain
from minimising $d_{KL}$. Assume that we have
already computed an interpolation $u(\vx)$ with $N$ grid points,
then the Kullback-Leibler distance $d_{KL}$ when evaluating the
function at an additional point $\vx_{N+1}$ is changed by
\begin{eqnarray}
 \nonumber \left|
d_{KL}^{\rm new} - d_{KL}^{\rm old} \right|\!\!\! &=&\!\!\!
\bigg| \int d^dx\, \LL(\vx) \bigg[\LLL(\vx) - u^{\rm new}(\vx) . \\ \nonumber
\!\!\!&&\!\!\!  -\LLL(\vx) + u^{\rm old}(\vx) \bigg] \bigg|\\ \nonumber 
\!\!\!&=&\!\!\! \left|
\int d^dx\, \LL(\vx) \left[u^{\rm old}(\vx) - u^{\rm new}(\vx) \right]
\right| \\ \nonumber 
\!\!\!&=&\!\!\! \left| \int d^dx\, \LL(\vx)
\left[\sum_{i=1}^{N} \alpha_{i} \varphi_{i}(\vx) - \sum_{i=1}^{N+1}
  \alpha_{i} \varphi_{i}(\vx) \right] \right| \\  
\!\!\!&=&\!\!\! \left|
\int d^dx\, \LL(\vx) \left[\alpha_{N+1} \varphi_{N+1}(\vx) \right]
\right|\,.
\label{refinement_compl}
\end{eqnarray}
If we refine the interpolation around the grid point that
contributed most to the Kullback-Leibler distance, we can hope to
converge towards the minimum of $d_{KL}$ fastest. In order to obtain a
suitable refinement criterion, we have to simplify the formula in
(\ref{refinement_compl}) considerably.
We thus assume the likelihood $\LL(\vx)$ as well as the basis
function $\varphi_{N+1}(\vx)$ to be locally constant on
$\varphi_{N+1}$'s support, obtaining
\begin{equation}
\left| 
d_{KL}^{\rm new} - d_{KL}^{\rm old} \right| \sim V_{N+1}
\LL(\vx_{N+1}) \,| \alpha_{N+1} |\,,
\label{refinement}
\end{equation}
where $V_{N+1}$ is the volume covered by the basis function $\varphi_{N+1}$
(\ie its support), and we have used $\varphi_{N+1}(\vx_{N+1}) = 1$.

With (\ref{refinement}), we have derived an estimation of the
contribution of a basis 
function to $d_{KL}$, which is a reasonable refinement criterion in
our setting. In addition to the surplus of the grid point, $|
\alpha_{N+1} |$, it takes into account the value of the likelihood
$\LL(\vx_{N+1})$ at the grid point, and the volume of the basis
function $V_{N+1}$.
The likelihood takes care of the fact that we would like to be more
accurate where the likelihood is higher. The regions of very low
likelihood are less interesting for us---the likelihood being already
very close to zero beyond a difference of about 20 log-likelihoods.  The
volume factor, on the other hand, prevents the interpolation to refine
too deeply (to very high grid levels) locally in the parameter
space. However, since this usually only takes effect after several
refinement steps, and as we have restricted the number of grid points,
we choose not to include the volume factor but rather to refine several
points at the same time, which addresses this issue in an alternative
way, and which will be discussed later on.
We further choose to introduce a temperature $T$ again, which allows
us to weight the likelihood with respect to the surplus and thus to
influence how much to refine close to the maximum. The refinement
criterion we used in this study is thus
\begin{equation}
\left(\frac{\LL(\vx_{\vl,\vi})}{\LL_\Max}\right)^{1/T} \,| \alpha_{\vl,\vi} |\,,
\label{eq:refinement_criteria_likeli}
\end{equation}
where we have divided the likelihood by its peak value, $\LL_\Max$,
(which we have already obtained determining the interpolation domain)
because the WMAP code returns the log-likelihood only up to a constant
offset, so that we do not know the correct normalisation of $\LL$.
For $T=1$, refinement takes place only very close to the maximum as
$\LL$ decays quickly; a temperature of $T=6$ showed to provide good
results within the whole domain of interest.

Refining only one grid point per refinement step often causes
adaptivity to get stuck in a single, special characteristic of the
function. Interpolating $\LLL$ with our choice of basis functions, all
grid points are likely to be created only in the direction where the
log-likelihood decays fastest, or around one of the local dips we will
address later on. Refining more than one grid point at the same time
helps to circumvent such effects, resulting in a broader refinement
scheme.

The Kullback-Leibler distance $d_{KL}$ can also be used to measure the
quality of our interpolation: The distance between the real likelihood
and or interpolation should be as small as possible. However,
as we have already mentioned above,
we do not know the normalisation of the WMAP likelihood
function. Therefore, $d_{KL}$ is not necessarily positive and
thus looses its property of being a useful measure of the `closeness'
of the two functions. We thus use a slightly modified version,
\begin{equation}
 {\widehat d}_{KL} \equiv \int d^dx\, \LL(\vx) |\LLL(\vx) - u(\vx)|\,,
\label{KL}
\end{equation}
as a measure of the quality of our interpolation,
instead of the actual Kullback-Leibler distance. 
It can be easily calculated from an MCMC with
  $T=1$ obtained for $\LL$, by simply averaging the
absolute errors $|\LLL(\vx_i) - u(\vx_i)|$ over all
points in the chains. Furthermore, we quote this value averaged over a chain of
$T=3$ (exploiting the interpolation domain better), which corresponds
to $\int d^dx\, \LL(\vx)^{\frac{1}{T}} |\LLL(\vx) - u(\vx)|$.

\begin{figure}
\centering
\includegraphics[angle=270,width=0.4\textwidth]{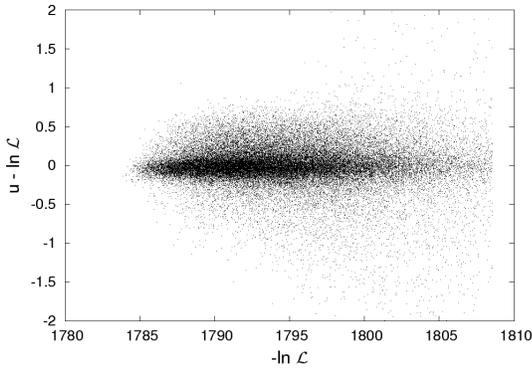}
\caption[Scatterplot, 7 dimensions, standard parameters]{Absolute
  error of the interpolation with respect to the real 
  log-likelihood in 7 dimensions for an interpolation with an
  adaptively refined sparse grid for standard parameters. 
} 
\label{fig:dirk_cosmos_scatter_7dim_adapt}
\end{figure}

Figure \ref{fig:dirk_cosmos_scatter_7dim_adapt} shows 
the scatterplot for an adaptively refined sparse grid in 7
dimensions. Starting from a regular grid of level 5, we refined 100
grid points each according to the refinement criterion
(\ref{eq:refinement_criteria_likeli}) with $T=6$. 
Needing only about as much grid points (78551) as for the regular
sparse grid of level 7, Tab.\ \ref{tab:error} shows that we obtain
results which are close to those of a regular sparse grid of level 8
with almost 4 times as many grid points. We provide the Mean Squared
Error (MSE) as well as ${\widehat d}_{KL}$ for both $T=1$ and $T=3$
chains for regular sparse grids of level 7 and 8, and for the
adaptively refined case. We also quote how many points exhibit an
absolute error larger than 1 or 0.25 for the $T=3$ chains. We do not
show the histograms of the adaptively refined model, as the histograms
for both the regular grid on level 8 and the adaptively refined one, are
very close to the histograms for normal parameters in 7 dimensions
shown in Figure \ref{fig:salsa_7dim_level6}.

\begin{table*}
\begin{centering}
\begin{tabular}{lcccccc}
\hline
 & err $> 1$, T=3 & err $> 0.25$, T=3 & MSE, T=1 & MSE, T=3 & $\widehat d_{KL}$,
T=1 & $\widehat d_{KL}$, T=3
\vspace{0.1cm}\\
level 7 & 4.2\% & 50.5\% & 0.087 & 0.532 & 0.256 & 0.354 \\
level 8 & 2.3\% & 19.3\% & 0.017 & 0.210 & 0.091 & 0.193 \\
adaptive& 1.8\% & 23.6\% & 0.027 & 0.202 & 0.110 & 0.204 \\
\hline
\end{tabular}
\caption[Comparison of errors for adaptive refinement]{Comparison of
  errors of regular sparse grids of level 7 and 
  level 8, respectively, and an adaptively refined sparse grid using
  approximately as many grid points as contained in the regular grid
  of level 7. Shown are the number of points with an absolute error
  larger than 1 or 0.25 in the $T=3$ chains, the MSE for 
  chains of $T=1$ and $T=3$, and ${\widehat d_{KL}}$, which denotes the absolute
  value of the error averaged over chains of $T=1$ and $T=3$.}
\label{tab:error}
\end{centering}
\end{table*}

We would like to mention, that, due to numerical problems, the current
version of {\small CMBEASY} produces local, unphysical and sometimes rather
high dips. This problem is already known and will be corrected in the
next release. For stochastic approaches, this is not a big problem,
though: The dips are local and just cause some noisy evaluations. But
it poses a problem for numerical approaches if a grid point hits a
dip. Then it can happen, that spending more grid points can even
deteriorate the results. For our regular grid in 6 dimensions using
the standard parameters, e.g., increasing the level from 7 to 8 caused a
higher overall error on the chain-data used for the histograms, as
especially two new basis functions close to the peak caused an error
of up to 12 of the log-likelihood for all evaluations affected by
those basis functions.

Fortunately, dips can be detected automatically due to the
hierarchical structure of the sparse grid and the smoothness of
$\LLL$, using a criterion that is once more based on the
surpluses. Furthermore, it is not a severe problem when using
adaptivity, as adaptivity localises the effects of the dips
automatically. One just has to take care not to spend too much grid
points to compensate for the dips.

The first adaptive results are promising, but there is still room for
a lot of improvement. Even better refinement criteria than those used
so far could be employed. Using not only piecewise linear functions, but
rather piecewise polynomials, and applying adaptivity in both the
mesh-width and the polynomial degree is very promising; especially the
extrapolation properties towards the boundary would be improved, and
less grid points would be needed to obtain the same accuracies.

\section{Conclusions}\label{sec:conclusions}

In this work, we have explored the utility of interpolating the WMAP
log-likelihood surface using sparse grids. We demonstrated that the
results are excellent and competitive to other approaches regarding
speed and accuracy, and we discussed advantages over fitting the
likelihood surface with polynomials \citep{pico, cmbfit} or neural
networks \citep{cosmonet}:

The interpolation based on sparse grids converges towards the exact
function in the limit of the grid level going to infinity. We can
therefore reach an arbitrary accuracy by simply increasing the amount
of work we spend. In the case of a polynomial fit, this is not
guaranteed since increasing the polynomial degree runs the risk of
becoming unstable.

In order to construct the sparse grid interpolation, we do not need to
sample a set of training points using MCMCs beforehand, since the
sampling points are determined by the sparse
grid structure which is 
given a priori. Once we have chosen the volume of interest,
the time for constructing the interpolation is dominated by
the evaluation of the likelihood function at the grid points. We do
not need additional training time as for neural networks
\citep{cosmonet}, for example. Constructing the
interpolation can thus be done 
almost arbitrarily in parallel, only limited by the computational
resources that are available.

The sparse grid technique is rather general and not restricted to
certain classes of functions. In particular, the choice of sampling
points and basis functions is not tailored to a 
single problem as for neural networks, where the topology of the
network has to be chosen problem-specifically (often in a heuristic
way). The sparse grid interpolation
technique as well as our extensions can therefore be readily applied
to other problems in astrophysics and cosmology, and will be useful in
further tasks, where an accurate interpolation of a function is needed.

The excellent performance of the sparse grid interpolation can be
further improved, leaving future research to do: It can be applied to
models with more than seven parameters by spending more computational
effort. Further modification of the basis functions, for example
allowing for a piecewise polynomial interpolation, promises better
convergence rates and higher accuracies.  Adaptive refinement schemes,
which take into account the characteristics of the interpolated
function, can be used to further increase the accuracy of the
interpolation, as we have already demonstrated for a first example in
this work.

\section*{Acknowledgements}

The authors would like to thank Georg Robbers for his help with
{\small CMBEASY} and Stefan Zimmer and the anonymous referee for
helpful discussions and comments. We acknowledge the use of {\small
  CMBEASY} and the WMAP likelihood code.

\bibliographystyle{mn2e}
\bibliography{bibl,dirk}

\begin{appendix}
\chapter{}

\section{Inversion of the parameter transformation}\label{app:sparse}
\begin{figure} 
  \centering
  \includegraphics[scale=0.6]{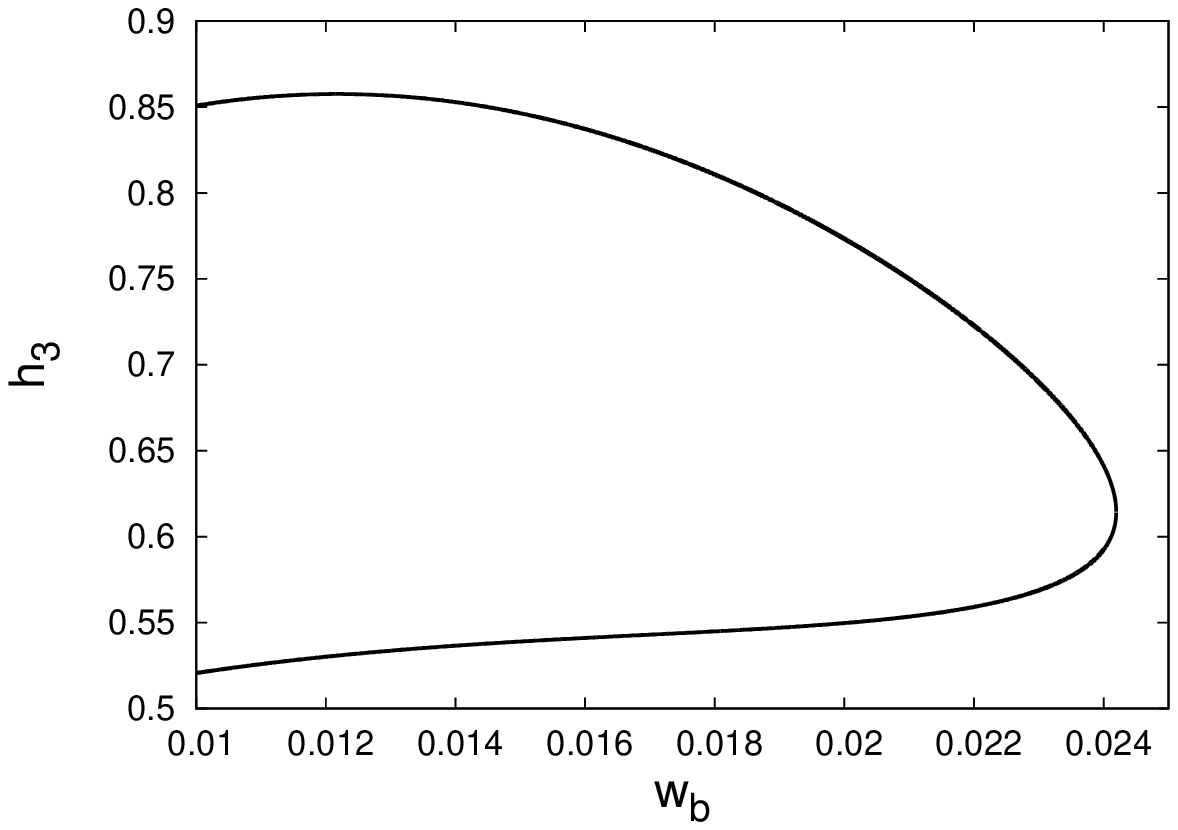}
  \caption{The two branches of $h_3$ versus $w_b$ for $h_2 = 0.45$.} 
  \label{fig:h3}
\end{figure}

In the following, we present a technique of inverting the parameter
transformation of Sec.\ \ref{sec:trafo} to compute the cosmological
parameters given the normal parameters.  The normal parameter $h_2$ in
terms of cosmological parameters is given by 
\bea \nonumber
h_2(w_m,w_b) \!\!&=&\!\!
0.0264\,w_b^{-0.762}\\
\!\!&&\!\!\exp\left(-0.476\left[\ln(25.5\,w_b+1.84\,w_m)\right]^2\right)\,.
\eea
We solve this equation for $w_m$ as a first step: 
\bea \nonumber
w_m(h_2,w_b)\!\!\!\! &=& \!\!\!\! \bigg(\exp\left\{\pm\left[ -\frac{1}{0.476}\ln\left(
  \frac{h_2}{0.0264}\,w_b^{0.762}\right)\right]^{1/2}\right\}\\
\!\!\!\! &&\!\!\!\! -25.5\,w_b \bigg) \frac{1}{1.84}\,.  
\label{wm} 
\eea 
Inconveniently, there exist
two different solutions for $w_m(h_2,w_b)$, which complicates the
inversion. We now substitute $w_m$ in $h_3(w_m,w_b)$ (\ref{h3}) for
(\ref{wm}) and thus obtain $h_3(h_2,w_b)$, which, of course, has two
solutions as well. An example of the two branches of $h_3(h_2,w_b)$ for
$h_2 = 0.45$ is depicted in Fig.\ \ref{fig:h3}.
We can compute the critical point where only one solution exists using
the condition 
\be
-\frac{1}{0.476}\ln\left(\frac{h_2}{0.0264}\,w_b^{0.762}\right) = 0\,,
\ee
as can be seen from (\ref{wm}). This condition gives us the following
formulae for the parameter values at the critical point:
\bea
w_{b,{\rm crit}}(h_2) &=& \left(\frac{0.0264}{h_2}\right)^{1/0.762}\,, \\
w_{m,{\rm crit}}(h_2) &=& \left(1-25.5\,w_{b,{\rm
    crit}}\right)\frac{1}{1.84}\,, \\ \nonumber
h_{3,{\rm crit}}(h_2) &=& 
2.17\left(1+\left(\frac{w_{b,{\rm crit}}}{0.044}\right)^2\right)^{-1} 
w_{m,{\rm crit}}^{0.59} \\
&&\left(1+1.63  
\left(1-\frac{w_{b,{\rm crit}}}{0.071}\right)w_{m,{\rm crit}}\right)^{-1}\,.
\eea
The two parameters $h_2$ and $h_3$ can now be inverted to
$w_m$ and $w_b$. For a given $h_2$, we express $h_3$ in terms of $h_2$ and
$w_b$, as described above. We then use $h_{3,{\rm crit}}(h_2)$ to
choose the upper branch of $h_3(h_2,w_b)$ if our given $h_3$ is bigger than
$h_{3,{\rm crit}}(h_2)$, and the lower branch if it is smaller. Using the
respective branch of $h_3(h_2,w_b)$, we search numerically in $w_b$ until
$h_3(h_2,w_b)$ matches the given $h_3$. Substituting that value of $w_b$
into equation (\ref{wm}), we readily obtain the value for $w_m$.

Now it is straightforward to compute the values for $n_s$ and $A_s$ from $t$ and
$A_*$. To obtain $h$ from $\Theta_s$, we follow the procedure suggested by
\cite{kosowsky}, expressing $\Theta_s$ in terms of $h$ in terms of $h$
and then searching in $h$ numerically until $\Theta_s(h)$ matches
the given value of $\Theta_s$.

\end{appendix}

\label{lastpage}

\end{document}